\documentclass[lettersize,journal]{IEEEtran}
\usepackage{amsmath,amsfonts}
\usepackage{algorithmic}
\usepackage{algorithm}
\usepackage{array}
\usepackage{textcomp}
\usepackage{stfloats}
\usepackage{url}
\usepackage{verbatim}
\usepackage{graphicx}
\usepackage{cite}
\usepackage{multirow}
\usepackage{makecell}
\usepackage{hyperref}
\usepackage{booktabs}

\usepackage{subcaption}
\usepackage{adjustbox}

\usepackage{color, colortbl}
\newcommand{\gc}{\cellcolor[gray]{0.915}}
\usepackage{xcolor}

\usepackage{amsthm} 
\newtheorem{theorem}{Theorem}

\graphicspath{{figures/}}

\begin{document}

\title{Editable-DeepSC: Reliable Cross-Modal Semantic \\ Communications for Facial Editing}

\author{Bin Chen, Wenbo Yu, Qinshan Zhang, Tianqu Zhuang, Hao Wu, Yong Jiang, and Shu-Tao Xia
\thanks{Bin Chen is with the School of Computer Science and Technology, Harbin Institute of Technology, Shenzhen, Guangdong 518055, China (e-mail: chenbin2021@hit.edu.cn).}
\thanks{Wenbo Yu, Qinshan Zhang, Tianqu Zhuang, and Hao Wu are with the Tsinghua Shenzhen International Graduate School, Tsinghua University, Shenzhen, Guangdong 518055, China (e-mail: wenbo.research@gmail.com; \{zhangqs24, zhuangtq23, wu-h22\}@mails.tsinghua.edu.cn).}
\thanks{Yong Jiang and Shu-Tao Xia are with the Tsinghua Shenzhen International Graduate School, Tsinghua University, and also with the Peng Cheng Laboratory, Shenzhen, Guangdong 518055, China (e-mail: \{jiangy, xiast\}@sz.tsinghua.edu.cn).}
\thanks{Corresponding author: Hao Wu (e-mail: wu-h22@mails.tsinghua.edu.cn).}
}



\maketitle

\begin{abstract}
Interactive computer vision (CV) plays a crucial role in various real-world applications, whose performance is highly dependent on communication networks. Nonetheless, the data-oriented characteristics of conventional communications often do not align with the special needs of interactive CV tasks. To alleviate this issue, the recently emerged semantic communications only transmit task-related semantic information and exhibit a promising landscape to address this problem. However, the communication challenges associated with Semantic Facial Editing, one of the most important interactive CV applications on social media, still remain largely unexplored. In this paper, we fill this gap by proposing Editable-DeepSC, a novel cross-modal semantic communication approach for facial editing. Firstly, we theoretically discuss different transmission schemes that separately handle communications and editings, and emphasize the necessity of Joint Editing-Channel Coding (JECC) via iterative attributes matching, which integrates editings into the communication chain to preserve more semantic mutual information. To compactly represent the high-dimensional data, we leverage inversion methods via pre-trained StyleGAN priors for semantic coding. To tackle the dynamic channel noise conditions, we propose SNR-aware channel coding via model fine-tuning. Extensive experiments indicate that Editable-DeepSC can achieve superior editings while significantly saving the transmission bandwidth, even under high-resolution and out-of-distribution (OOD) settings.
\end{abstract}



\section{Introduction}

In the era of rapid technological advancement, interactive computer vision (CV) plays a pivotal role in various real-world applications, such as digital manufacturing \cite{chryssolouris2009digital}, telemedicine \cite{alenoghena2023telemedicine}, robotics \cite{soori2023artificial}, and metaverse \cite{wang2023survey}. The performance of interactive CV tasks is heavily dependent on the underlying communication networks. These traditional communication systems are primarily designed for data-oriented purpose and focus on metrics that measure the data transmission instead of tasks execution performance. In other words, the original data is expected to be completely recovered at the receiver side regardless of the users' usage about it. Nonetheless, this data-oriented design principle is not always suitable for the specific needs of interactive CV tasks, as different tasks at the receiver side would require different types of extracted semantic information from the original data, which should be encoded and transmitted with different importance. Reconstructing every part of the original data equally will undoubtedly waste the limited transmission bandwidth.

\begin{figure}
\includegraphics[width=\linewidth]{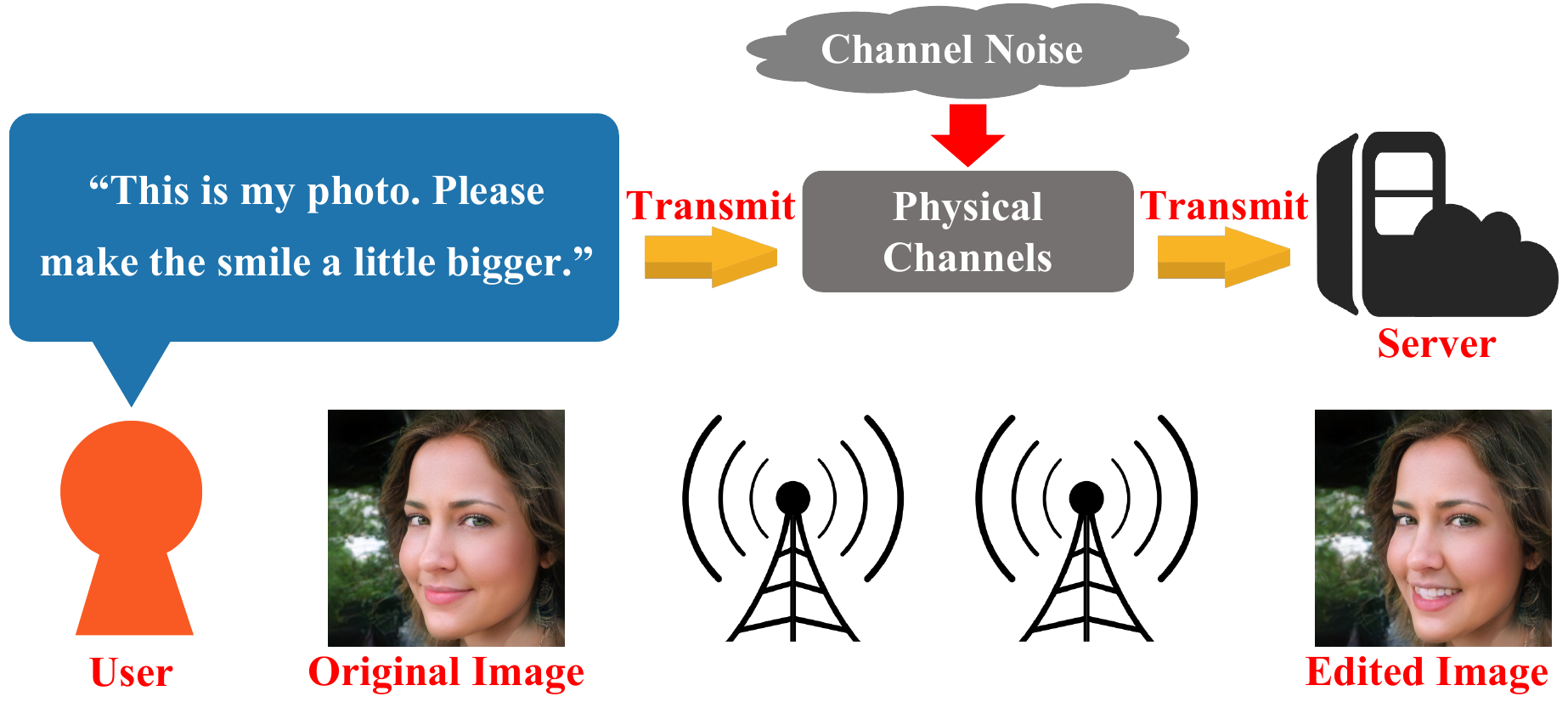}
\caption{Illustration of the dynamic semantic facial editing scenarios. During the transmission, users may wish to flexibly edit the original multimedia data according to their personal needs in a conversational and interactive way.}
\label{fig:dynamic_editing_scenarios}
\end{figure}

To alleviate this issue, the recently emerged semantic communications \cite{zhang2022deep, jiang2022wireless, wang2022wireless, xie2021task, wang2022distributed} (also known as task-oriented communications) only transmit the specific semantic information suitable for interactive CV tasks and exhibit a promising landscape to address this problem. By bridging the gap between bit-level transmission and task-oriented requirements, semantic communications can exploit much more of the scarce bandwidth to transmit task-related information, thus realizing better interactive CV tasks execution performance. On the other hand, Semantic Facial Editing \cite{yang2018learning, olszewski2020intuitive, wang2018every, shen2020interpreting, zhuang2021enjoy, jiang2023talk}, which stands out as one of the most important interactive CV tasks for user interaction and personalization, encounters various communication challenges when promoting its real-world deployment. As shown in Figure \ref{fig:dynamic_editing_scenarios}, in many famous social platforms (e.g., Facebook, Instagram), before transmitting their own photos to the remote servers, users may wish to flexibly edit the original data according to their personal needs, such as adding smile to a face or altering the transparency of eyeglasses. Furthermore, such personal requirements can be fulfilled by providing dialogues to enable a more conversational and interactive experience. A crucial question still remains largely unexplored in existing works: \emph{How to better transmit the facial semantics while also dynamically editing them according to the users' needs?}

In this paper, we fill this gap by proposing Editable-DeepSC, a novel cross-modal semantic communication approach for facial editing. Editable-DeepSC takes cross-modal text-image pairs as the inputs and transmits the edited semantic information of facial images based on textual instructions. For such dynamic facial editing and communication scenarios, Editable-DeepSC faces the following challenges:

\begin{enumerate}
    \item[\emph{$Q_1$}:] \emph{How to better compress the original data, so that satisfying editing effects can be achieved with less transmission bandwidth consumption?}
    \item[\emph{$Q_2$}:] \emph{How to better realize a trade-off between editing effects and fidelity, so that targeted semantics are edited while other untargeted semantics still remain unaffected?}
    \item[\emph{$Q_3$}:] \emph{How to better tackle the dynamically varying channel conditions, so that the model can generalize well to different circumstances of communication capabilities?}
\end{enumerate}

To address the above challenges, we propose Editable-DeepSC, a novel framework that integrates facial editing directly into the semantic communication chain. Unlike conventional separate schemes (where editing is performed either before or after transmission) that introduce redundant data processing procedures and risk losing semantic mutual information, our approach employs a joint optimization strategy to preserve semantic fidelity and enhance editing quality.

Editable-DeepSC operates through three cohesive stages. First, it leverages generative priors for efficient semantic coding, dramatically compressing high-dimensional facial images into low-dimensional representations to tackle \emph{$Q_1$}. Second, it introduces a Joint Editing-Channel Coding (JECC) mechanism that simultaneously performs semantic editing and ensures robustness against channel noise, addressing the fine-grained control requirements of \emph{$Q_2$}. Finally, an SNR-aware adaptation mechanism enables the model to efficiently generalize across diverse channel conditions by fine-tuning only a minimal set of parameters, effectively solving \emph{$Q_3$}.

To verify the effectiveness of Editable-DeepSC, we conduct extensive experiments under various levels of SNRs and compare it with the baselines that separately handle communications and editings, where both editings after the channel transfer and before the channel transfer are evaluated. We also consider more rigorous and realistic simulation settings where the transmitter side users may hold high-resolution images or out-of-distribution (OOD) images. Numerous results indicate that Editable-DeepSC achieves superior editing effects compared to the baselines while significantly saving the transmission bandwidth. For instance, in high-resolution settings, Editable-DeepSC only consumes less than $2\%$ of the baseline methods' bandwidth, but still achieves the best editing performance both quantitatively and qualitatively. Furthermore, in the fine-tuning stage, Editable-DeepSC only adjusts $2.65\%$ of the total parameters, but still considerably improves the editing effects under low SNRs and protects the edited semantics from channel noise corruptions.

To the best of our knowledge, we are the first to systematically and comprehensively investigate the communication problems of dynamic facial editing in this research field. Our main contributions are as follows:

\begin{itemize}
\item We theoretically analyze different transmission schemes that separately handle communications and editings, encompassing both transmitter-side (pre-channel) and receiver-side (post-channel) editings. We discover that they actually increase the data processing procedures and are more likely to lose semantic mutual information, undermining efficiency and ultimately degrading received image quality. In light of this, we emphasize the necessity to integrate editings into the communication chain, jointly optimize these two aspects, and minimize the data processing procedures to preserve more semantic mutual information.
\item We propose semantic coding via Generative Adversarial Networks (GAN) inversion. By leveraging the abundant generative priors within the pre-trained StyleGAN, we encode the high-dimensional input images into the GAN latent space and map them to low-dimensional representations. Thus, we compactly represent the high-dimensional input data and considerably decrease the Channel Bandwidth Ratio (CBR) required to transmit the edited facial semantics.
\item We introduce Joint Editing-Channel Coding (JECC) via iterative attributes matching. By employing low-complexity Fully Connected Layers and an attribute-matching loop, we iteratively exert minor modifications on the results of semantic coding to realize semantic editing and channel coding at the same time. Thus, we reduce the data processing procedures for semantic mutual information preservation and precisely perform fine-grained editings only on the targeted semantics, without influencing other untargeted ones.
\item We present SNR-aware channel coding via model fine-tuning. We introduce two lightweight trainable adapters (which do not change the shapes of inputs and outputs within the JECC codecs) to capture the distribution of new noise conditions. When fine-tuning the models, only the parameters of these two adapters are adjusted to learn the characteristics of new transmission environments. Meanwhile, the rest of the parameters are frozen to avoid forgetting the previously learned priors due to the fine-tuning. This helps the model generalize well to the varying communication capabilities via minimal parameters fine-tuning.
\item Extensive experiments demonstrate that Editable-DeepSC achieves superior performance compared to existing methods on editing effects and transmission efficiency, even under more practical settings with high-resolution and out-of-distribution (OOD) images.
\end{itemize}

The rest of this paper is organized as follows. Section \ref{related_works} illustrates the related works. Section \ref{system_model} introduces the system model. Section \ref{method} elaborates on the implementation of Editable-DeepSC. Section \ref{experiments} provides the experiment results. Section \ref{conclusion} concludes this work.

\section{Related Work}
\label{related_works}

\subsection{Semantic Facial Editing}

Semantic Facial Editing aims to provide personalized facial manipulations specified by the users, and has emerged as one of the most important interactive CV tasks for its wide range of applications, including virtual content creation \cite{wang2020vr}, augmented reality (AR) \cite{arena2022overview}, and online conferencing \cite{ryu2021performance}. This technology enables users to make intuitive edits to facial attributes such as expression, age, hairstyle, offering both creative flexibility and enhanced user experiences.

In the early stage of this area, many methods \cite{olszewski2020intuitive, wang2018every} focus on editing specific attributes, and rely heavily on extra manually provided knowledge such as landmarks. Recently, latent space based facial editing methods \cite{shen2020interpreting, zhuang2021enjoy, jiang2023talk} are proposed due to the advancement of generative models like StyleGAN \cite{karras2019style}. These methods typically strive to find semantically meaningful directions in the latent space of pre-trained generative models, so that shifting along the latent space would be able to achieve the desired editing in the vision space. InterFaceGAN \cite{shen2020interpreting} finds a hyperplane in the latent space to disentangle the facial semantics, and uses the normal vector of the hyperplane as the editing direction. Zhuang \emph{et al.} \cite{zhuang2021enjoy} proposed to learn a mapping network to generate identity-specific directions within the latent space. However, both the two methods \cite{shen2020interpreting, zhuang2021enjoy} neglect language guidance and lack conversational interaction. To alleviate this issue, Jiang \emph{et al.} \cite{jiang2023talk} proposed Talk2Edit, which enables customized editings via human language feedback.

\subsection{Task-Oriented Semantic Communications}

Semantic communications prioritize the transmission of task-related semantic information rather than the original data with enhanced communication efficiency, and have showed great potential across various CV tasks.

\textbf{Single-modal semantic communications for image tasks.} Zhang \emph{et al.} \cite{zhang2022deep} proposed a DNN-based image semantic communication system, which only transmits the gradients back to the transmitter during training to protect the receiver's privacy of downstream tasks. Han \emph{et al.} \cite{han2025scsc} proposed Standards-Compatible Semantic Communication (SCSC), an end-to-end image transmission approach with learnable preprocessing-empowered and precoder/combiner-enhanced modules for MIMO channels. Huang \emph{et al.} \cite{huang2026robust} proposed a wavelet transform frequency-domain attention feature module, which employs multilevel wavelet decomposition to effectively separate semantic structures from high-frequency adversarial noise.

\textbf{Single-modal semantic communications for video tasks.} Jiang \emph{et al.} \cite{jiang2022wireless} proposed a video semantic coding network based on keypoints transmission, and considerably reduced resources consumption without losing the main details. Wang \emph{et al.} \cite{wang2022wireless} exploited nonlinear transform and conditional coding architecture to adaptively extract semantic features across video frames, and surpassed traditional wireless video coded transmission schemes. Li \emph{et al.} \cite{li2024video} proposed a video semantic communication system based on major object extraction and contextual video encoding, which can achieve efficient video transmission for massive communication.

\textbf{Semantic communications for multi-modal/cross-modal CV tasks.} Xie \emph{et al.} \cite{xie2021task} proposed a multi-modal semantic communication system for Visual Question Answering (VQA) tasks, which utilizes MAC networks \cite{hudson2018compositional} to deal with the interconnected cross-modal data and generate the answers to VQA tasks. Jiang \emph{et al.} \cite{jiang2025visual} proposed a novel Vision-Language Model-based Cross-modal Semantic Communication (VLM-CSC) system, which consists of three innovative components: Cross-modal Knowledge Base (CKB), Memory-assisted Encoder and Decoder (MED), and Noise Attention Module (NAM). VLM-CSC effectively addresses the challenges in dynamic environments such as low information density and catastrophic forgetting. Tian \emph{et al.} \cite{tian2025synchronous} proposed a synchronous multi-modal semantic communication System with packet-level coding (SyncSC) for facial video and speech transmission, which reduces transmission overhead while ensuring high-quality synchronous transmission of video and speech over the packet loss network.

However, to the best of our knowledge, the cross-modal semantic communication challenges under dynamic facial editing scenarios have not yet been systematically explored up to now. In this paper, we aim to fill this gap and comprehensively address these problems.


\section{System Model}
\label{system_model}

\begin{figure*}[t]
\centerline{\includegraphics[width=0.75\textwidth]{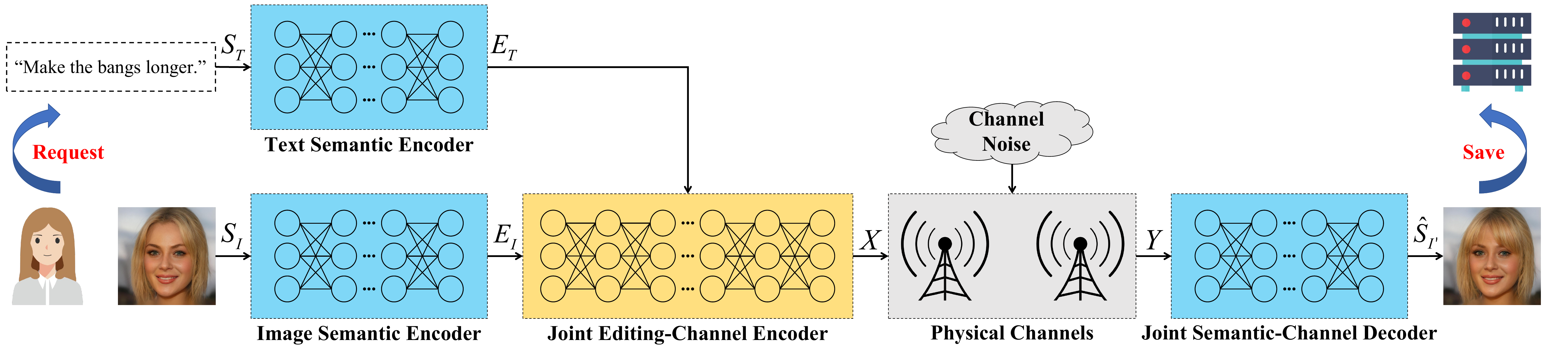}}
\caption{Overview of the proposed framework. Editable-DeepSC mainly consists of the Text Semantic Encoder, the Image Semantic Encoder, the Joint Editing-Channel Encoder, and the Joint Semantic-Channel Decoder, where channel noise corruptions from the real world are also considered.}
\label{fig:overview}
\end{figure*}

As shown in Figure \ref{fig:overview}, our proposed Editable-DeepSC mainly consists of the Text Semantic Encoder, the Image Semantic Encoder, the Joint Editing-Channel Encoder, and the Joint Semantic-Channel Decoder, where channel noise corruptions from the real world are also taken into consideration.

\subsection{Semantic Transmitter}

The input image $S_{I} \in \mathbb{R}^{1 \times C \times W \times H}$ (where $H$, $W$, and $C$ represent the image's height, width, and color channels) is first transformed by the Image Semantic Encoder into low-dimensional representations for data compression, i.e.,
\begin{equation}
    E_{I} = \mathcal{SE_{I}}(S_{I};\alpha),
\end{equation}

\noindent where $E_{I}\in\mathbb{R}^{1 \times L_1}$ (with $L_1$ being the dimension of visual features) is the extracted image semantic information and $\mathcal{SE_{I}}(\cdot;\alpha)$ denotes the Image Semantic Encoder with the parameters $\alpha$. Similarly, the corresponding instructive sentence $S_{T} \in \mathbb{R}^{1 \times L_2 \times L_3}$ (where $L_2$ is the text length and $L_3$ is the dimension of textual embeddings) is also encoded by the Text Semantic Encoder $\mathcal{SE_{T}}(\cdot;\beta)$ with the parameters $\beta$ to acquire the textual semantic information $E_{T}\in\mathbb{R}^{1 \times L_4}$ (with $L_4$ being the dimension of textual features), which is highly relevant to the anticipated editing:
\begin{equation}
    E_{T} = \mathcal{SE_{T}}(S_{T};\beta).
\end{equation}

Next, $E_{I}$ and $E_{T}$ are sent into the Joint Editing-Channel Encoder $\mathcal{JECE}(\cdot, \cdot;\gamma)$ with the parameters $\gamma$, which will apply semantic editing as well as channel coding to the transmitted information in the meantime:
\begin{equation}
    X = \mathcal{JECE}(E_{I}, E_{T};\gamma).
\end{equation}

The output of $\mathcal{JECE}(\cdot, \cdot;\gamma)$, i.e., $X\in\mathbb{C}^{1 \times k}$ (where $k$ is the length of $X$), is then transmitted through the physical channels. We enforce an average transmission power constraint on $X$ to ensure that it does not exceed the limits of current communication capabilities:
\begin{equation}
    \frac{1}{k} \cdot ||X||^{2}_{2} \leq P_{max},
\end{equation}

\noindent where $P_{max}$ is the power constraint.

Following the previous works \cite{erdemir2023generative, yilmaz2023distributed, tung2023deep} in this research field, the Channel Bandwidth Ratio (CBR) can be defined as:
\begin{equation}\label{eq:CBR}
    \rho = \frac{k}{H \times W \times C}.
\end{equation}

\noindent Smaller $\rho$ indicates better compression. In our framework, the GAN inversion process (detailed in Section \ref{sec:gan_inversion}) plays a central role in achieving a low CBR. It compresses the high-dimensional input image $S_I \in \mathbb{R}^{1 \times C \times W \times H}$ into a low-dimensional latent representation $E_I\in\mathbb{R}^{1 \times L_1}$. This latent vector $E_I$ is further processed by the JECC encoder to constitute the signal $X\in\mathbb{C}^{1 \times k}$ for transmission, whose length $k$ is also very small due to the previous compression step of GAN inversion, consequently resulting in smaller $\rho$.

\subsection{Semantic Receiver}

The transmitted signals are usually corrupted by the channel noise. Consequently, only the disrupted forms of the edited semantic information can be detected at the receiver side, i.e.,
\begin{equation}\label{eq:channel_transfer}
    Y = h \ast X + N,
\end{equation}

\noindent where $Y \in \mathbb{C}^{1 \times k}$ represents the received signal, $h \in \mathbb{C}$ denotes the channel coefficients, and $N\in \mathbb{C}^{1 \times k}$ denotes the Gaussian noise, whose elements (including the real and imaginary parts) are independent of each other and have the same mean and variance. Under simplified channel models such as the Additive White Gaussian Noise (AWGN) channel, $h$ is typically a scalar constant (e.g., $h=1$, representing no fading). In more general flat-fading channel models, $h$ is a complex random variable following a certain distribution (e.g., Rayleigh, Rician) that characterizes the attenuation and phase shift introduced by the wireless channel. For simplicity, we consider $h=1$ following the previous methods \cite{yilmaz2023distributed, wang2022distributed}.

Finally, $Y$ is mapped back to the high-dimensional vision domain to get the edited image $\hat{S}_{I^{\prime}} \in \mathbb{R}^{1 \times C \times W \times H}$ required by the transmitter:
\begin{equation}
    \hat{S}_{I^{\prime}} = \mathcal{JSCD}(Y;\theta),
\end{equation}

\noindent where $\mathcal{JSCD}(\cdot;\theta)$ indicates the Joint Semantic-Channel Decoder with the parameters $\theta$. It first acts as a channel decoder, where its internal components (e.g., lightweight adapters) work to correct the errors introduced by the noisy channel, effectively denoising the received signal $Y$. Subsequently, it functions as a semantic decoder and maps the partially corrected latent representations back to the high-dimensional image space, reconstructing the final edited image $\hat{S}_{I'}$. This integrated process ensures that semantic recovery and channel error correction are jointly optimized, rather than being treated separately, which will reduce the semantic errors caused by the channel noise. Moreover, since only the edited images are ultimately required and there is no need to reconstruct the instructive sentences, the Text Semantic Decoder is unnecessary in our model and thus not considered.

\subsection{Theoretical Rationality}
\label{sec:theoretical_analysis}
To justify the rationality of our proposed Editable-DeepSC, we consider the alternative separate schemes below and theoretically demonstrate the superiority of our approach. Specifically, based on the sequence of communications and editings, we consider the following separate schemes:
\begin{enumerate}
    \item[\emph{$M_1$}:] \emph{(editings after the channel transfer) The transmitter sends the images, and then the receiver edits them.}
    \item[\emph{$M_2$}:] \emph{(editings before the channel transfer) The transmitter edits the images, and then sends them to the receiver.}
\end{enumerate}

For the encoding and decoding steps of DNN as well as the channel transfer steps, the outputs of each stage are largely dependent on the current inputs instead of the previous outputs, which approximately satisfies the Markov property. Thus, as suggested in many previous works \cite{tishby2015deep, saxe2018information, shao2021learning}, we approximate the above encoding, decoding, and channel transfer steps as a Markov chain $K_0 \rightarrow K_1 \rightarrow \cdot\cdot\cdot \rightarrow K_n$, where $K_0$ indicates the distribution of original images at the transmitter side, $n$ is the total number of data processing procedures, $K_n$ indicates the distribution of ultimate edited images at the receiver side, and $K_s \; (0 < s < n)$ indicates the distribution of a series of intermediate representations during the above steps. For any two transitions $U \rightarrow V \rightarrow W$ from the whole Markov chain, we have the Data Processing Inequality (DPI) \cite{cover1999elements} as demonstrated in Theorem \ref{DPI}.

\begin{theorem}
\textbf{(Data Processing Inequality)} If $U \rightarrow V \rightarrow W$, then $\mathcal{I}(U;V) \geq \mathcal{I}(U;W)$.
\label{DPI}
\end{theorem}

By applying the DPI to the whole Markov chain $K_0 \rightarrow K_1 \rightarrow \cdot\cdot\cdot \rightarrow K_n$ as pointed out in \cite{tishby2015deep}, we have $\mathcal{I}(K_0; K_1) \geq \mathcal{I}(K_0; K_2) \geq \cdot\cdot\cdot \geq \mathcal{I}(K_0; K_n)$. This implies that as the data processing procedures increase, the semantic mutual information about the original distribution $K_0$ carried by the intermediate distribution $K_s$ will decrease and cannot be recovered in the subsequent steps. Therefore, the aforementioned separate schemes \emph{$M_1$} and \emph{$M_2$} that have more encoding and decoding steps are more likely to lose semantic mutual information, which will eventually limit the quality of edited images. On the contrary, Editable-DeepSC integrates editings into the communication chain to decrease the data processing procedures, and will thus preserve more semantic mutual information with enhanced tasks performance.


\section{Method}
\label{method}

\begin{figure*}[t]
\centerline{\includegraphics[width=0.85\textwidth]{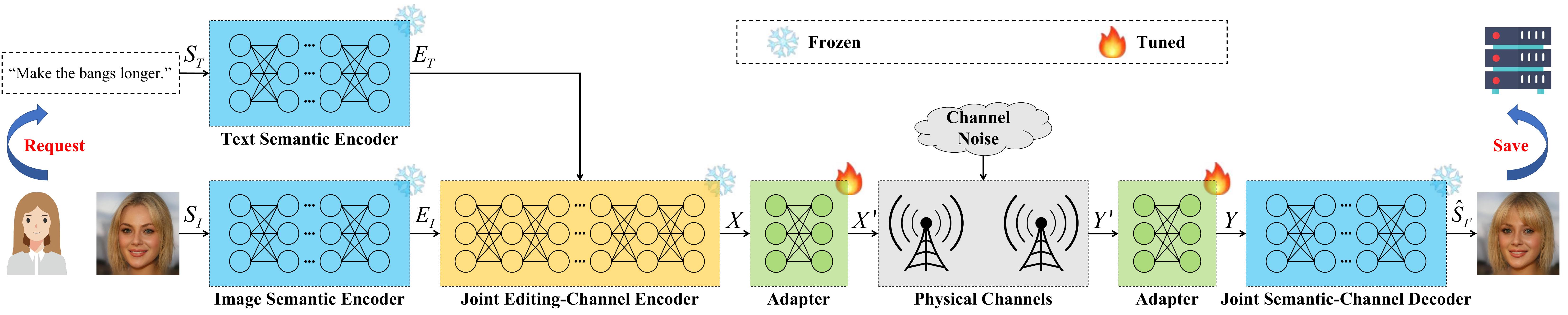}}
\caption{Illustration of our SNR-aware channel coding based on model fine-tuning. We introduce two lightweight trainable adapters that do not change the shapes of inputs and outputs to the Joint Editing-Channel Encoder and the Joint Semantic-Channel Decoder. When fine-tuning the models, only the parameters of adapters are adjusted to capture the distribution of new noise conditions, and the rest of the parameters are frozen to avoid forgetting the previously learned priors from Section \ref{sec:jecc}.}
\label{fig:overview_ft}
\end{figure*}

This section elaborates the proposed Editable-DeepSC framework, whose overall pipeline and corresponding three-stage data flow are illustrated in Figure \ref{fig:overview_ft}. The framework operates through three cohesive stages: (1) Semantic Coding via Pre-trained GAN Inversion; (2) Joint Editing-Channel Coding (JECC) via Iterative Attributes Matching; (3) SNR-aware Channel Coding via Model Fine-tuning.

In the first stage, the input facial image $S_I$ and textual instruction $S_T$ are processed by the Image Semantic Encoder $\mathcal{SE}_I(\cdot;\alpha)$ and Text Semantic Encoder $\mathcal{SE}_T(\cdot;\beta)$, to extract compact latent representations $E_I$ and $E_T$ via GAN inversion. These representations then flow into the second stage, where the core JECC mechanism iteratively refines $E_I$ guided by $E_T$ to produce the transmitted signal $X$, simultaneously achieving semantic editing and channel coding. After transmission through a noisy channel $Y = h \ast X + N$, the third stage employs SNR-aware fine-tuning, where lightweight adapters within $\mathcal{JECE}(\cdot, \cdot;\gamma)$ and $\mathcal{JSCD}(\cdot;\theta)$ are selectively updated to adapt to varying channel conditions, culminating in the generation of the final edited image $\hat{S}_{I'}$ by $\mathcal{JSCD}(\cdot;\theta)$. This integrated pipeline ensures precise execution of editings while maintaining robustness against channel noise, significantly reducing redundant data processing procedures compared to separate approaches.

\subsection{Semantic Coding via Pre-trained GAN Inversion}
\label{sec:gan_inversion}

We leverage the Generative Adversarial Networks (GAN) inversion methods \cite{goodfellow2020generative, xia2022gan} based on StyleGAN \cite{karras2019style} priors to design the Image Semantic Encoder $\mathcal{SE_{I}}(\cdot;\alpha)$, as there is abundant prior knowledge within the pre-trained GAN to reduce the representation dimension. To be specific, given an input image $S_I$, a random vector $z$ is first initialized in the StyleGAN latent space. The random vector $z$ is then fed into the pre-trained StyleGAN Generator $G(\cdot)$ to generate an initial image $I_g=G(z)$. The distortion $J(z)$ between $S_I$ and $I_g$ is calculated as:
\begin{equation}
    J(z) = \lambda_{inv} \cdot MSE(S_I, I_g) + LPIPS(S_I, I_g).
\end{equation}

\noindent The distortion $J(z)$ is measured from both the pixel level and the perceptual level, balanced by the hyper-parameter $\lambda_{inv}$. In this paper, we adopt the MSE loss for the pixel level distortion and the LPIPS loss \cite{zhang2018unreasonable} for the perceptual level distortion. To obtain the latent vector that perfectly suits the original image, we can minimize $J(z)$ and update $z$ by iteratively performing the gradient descent until convergence:
\begin{equation} \label{eq:inversion}
    z^{(t)} = z^{(t - 1)} - \eta_{inv} \cdot \nabla_{z^{(t - 1)}} J(z^{(t - 1)}),
\end{equation}

\noindent where $r_{inv}$ is the number of total inversion iterations, $z^{(0)}$ denotes the randomly initialized latent vector, $z^{(t)} \; (1 \leq t \leq r_{inv})$ denotes the outcome after the $t$-th iteration, and $\eta_{inv}$ is the inversion learning rate. The learning rate $\eta_{inv}$ is set to $0.10$ to ensure stable convergence, and the maximum number of iterations $r_{inv}$ is set to $110$ based on a trade-off between reconstruction quality and computational cost, as analyzed in Section \ref{sec:further_analysis}. The optimization process terminates when either the maximum iteration count $r_{inv}$ is reached or the relative change in the distortion $J(z)$ between consecutive iterations falls below a threshold $\epsilon = 1 \times 10^{-5}$, i.e., $|J(z^{(t)}) - J(z^{(t-1)})| / |J(z^{(t-1)})| < \epsilon$. Upon completing the optimization of (\ref{eq:inversion}), we will obtain the output of semantic coding, which is denoted as $E_{I} = z^{(r_{inv})}$. Note that the dimension of $z$ is often much smaller than the dimension of $S_I$, hence we can greatly compress the original data and reduce the Channel Bandwidth Ratio (CBR) defined in (\ref{eq:CBR}).

As for the Joint Semantic-Channel Decoder $\mathcal{JSCD}(\cdot;\theta)$, we first convert $Y$ back to a real-valued latent representation, which is then sent to another pre-trained StyleGAN Generator for semantic decoding to map the compressed data back to high-dimensional images. Besides, we additionally introduce lightweight trainable parameters for channel decoding to reduce the semantic errors caused by the channel noise, which we will detailedly discuss in Section \ref{sec:model_fine_tuning}.

\subsection{Joint Editing-Channel Coding (JECC) via Iterative Attributes Matching}
\label{sec:jecc}

\textbf{Editing Attributes.} Firstly, we need to quantitatively measure the attributes for editing (e.g., bangs, eyeglasses). We introduce an attribute predictor $P(\cdot)$ pre-trained on the CelebA-Dialog dataset \cite{jiang2021talk}, which consists of detailed manual annotations regarding the attributes of each training image. We send the original image to $P(\cdot)$ and obtain its output $\{a_1, a_2, ..., a_i, ..., a_m\}$, where the value of $a_i$ indicates the degree of each attribute (e.g., the length of the bangs, the transparency of eyeglasses), $m$ is the total number of attributes, and $u$ is the same maximum degree value for each attribute.

\textbf{Editing Semantic Information Extraction.} We utilize the Long Short-Term Memory (LSTM) \cite{hochreiter1997long} network to implement the Text Semantic Encoder $\mathcal{SE_{T}}(\cdot;\beta)$, which can capture the dependencies between various parts of the input texts. We train $\mathcal{SE_{T}}(\cdot;\beta)$ on the CelebA-Dialog dataset \cite{jiang2021talk} using the cross-entropy loss, where the training texts are manually annotated with their actual semantic information (i.e., the index of targeted attribute $i_{tar} \in \{1, 2, ..., m\}$, the direction of editing $d_{tar} \in \{-1, +1\}$, and the degree of modification $\Delta_{tar} \in \{0, 1, ..., u\}$) as the labels. In the test time, the original texts are tokenized into words or subwords according to the dictionary and then sent to $\mathcal{SE_{T}}(\cdot;\beta)$. The ultimate text encodings $E_T$ will contain the aforementioned semantic information needed for editing: $i_{tar}$, $d_{tar}$ and $\Delta_{tar}$.

\textbf{Joint Editing-Channel Coding (JECC).} Inspired by the fact that DeepJSCC \cite{bourtsoulatze2019deep} combines source coding and channel coding to jointly optimize data compression and error correction, we implement the Joint Editing-Channel Encoder $\mathcal{JECE}(\cdot, \cdot;\gamma)$ by Fully Connected Layers $\mathcal{M}(\cdot;\gamma)$ to jointly optimize editings and communications for reduced data processing procedures. The Joint Editing-Channel Encoder $\mathcal{JECE}(\cdot, \cdot;\gamma)$ will iteratively exert minor modifications on the previous results of semantic coding to realize semantic editing and channel coding at the same time. Note that $\mathcal{M}(\cdot;\gamma)$ only takes $E_I$ as the input, while $E_T$, the other input of $\mathcal{JECE}(\cdot, \cdot;\gamma)$, is utilized to obtain $i_{tar}$, $d_{tar}$ and $\Delta_{tar}$ that guide the semantic editing. In the test time, we first compute the expected degree of targeted attribute by $\{a_1, a_2, ..., a_i, ..., a_m\}$, $i_{tar}$, $d_{tar}$, and $\Delta_{tar}$:
\begin{equation}
    a_{exp} = min(max(a_{i_{tar}} + \Delta_{tar} \cdot d_{tar}, 0), u),
\end{equation}

\noindent where the minimum function $min(\cdot, \cdot)$ and the maximum function $max(\cdot, \cdot)$ are adopted to ensure that $0 \leq a_{exp} \leq u$. Then, we iteratively exert minor modifications on $E_I$:
\begin{equation} \label{eq:jecc}
    E_{I^{\prime}}^{(t)} = E_{I^{\prime}}^{(t - 1)} + d_{tar} \cdot \mathcal{M}(E_{I^{\prime}}^{(t - 1)};\gamma),
\end{equation}

\noindent where $r_{edit}$ is the maximum number of total editing iterations, $E_{I^{\prime}}^{(0)}$ is the same as $E_I$, and $E_{I^{\prime}}^{(t)} \; (1 \leq t \leq r_{edit})$ denotes the outcome after the $t$-th iteration. After each iteration of modification, $E_{I^{\prime}}^{(t)}$ is delivered to the StyleGAN Generator $G(\cdot)$ to reconstruct an intermediate image, which will be sent to the pre-trained attribute predictor $P(\cdot)$ defined in \cite{jiang2021talk} to check whether the anticipated requirement is satisfied. The predicted degree of the intermediate image calculated by $P(\cdot)$ will be compared with the expected degree $a_{exp}$. Once they match with each other, the iteration will stop and $E_{I^{\prime}}^{(t)}$ will be used to form the complex-valued signal $X$ (e.g., by treating the first and second halves as the real and imaginary parts) for transmission. Otherwise, $\mathcal{M}(\cdot;\gamma)$ will continue to make small movements on $E_{I^{\prime}}^{(t)}$ until the target is reached or the maximum number $r_{edit}$ is exceeded.

\textbf{Training Strategy of the JECC Codecs.} To train $\mathcal{M}(\cdot;\gamma)$ for the $i$-th attribute, we set the expected degree of the $i$-th attribute to be $(a_i + 1)$ so that the model can learn how to perform fine-grained editings on each attribute. Note that the situation for $(a_i - 1)$ is completely symmetrical when $d_{tar}$ is set to $-1$. The predictor loss $\mathcal{L}_{pred}$ can be calculated by adopting the cross-entropy loss:
\begin{equation}
    \mathcal{L}_{pred} = - \sum_{i = 1}^{m} \sum_{j = 0}^{u} b_{ij} \cdot log(p_{ij}),
\end{equation}

\noindent where $b_{ij} \in \{0, 1\}$ is the one-hot representation of the expected attribute degrees, and $p_{ij}$ is the output of the pre-trained attribute predictor $P(\cdot)$ on the edited image. We also leverage the identity preservation loss $\mathcal{L}_{id}$ to better keep the original facial identity. We use ready-made facial recognition model to extract the features. The extracted features for facial recognition should be as similar as possible:
\begin{equation}
    \mathcal{L}_{id} = || F(I_a) - F(I_b) ||_{1},
\end{equation}

\noindent where $I_a, I_b$ are the images after and before the editing, and $F(\cdot)$ is the pre-trained face recognition model \cite{deng2019arcface}. The discriminator loss $\mathcal{L}_{disc}$ is also adopted to ensure the fidelity on the generated images:
\begin{equation}
    \mathcal{L}_{disc} = - D(I_a),
\end{equation}

\noindent where $D(\cdot)$ is the output of the pre-trained StyleGAN \cite{karras2019style} discriminator. Finally, the total loss can be calculated as:
\begin{equation}
    \mathcal{L} = \lambda_{pred} \cdot \mathcal{L}_{pred} + \lambda_{id} \cdot \mathcal{L}_{id} + \lambda_{disc} \cdot \mathcal{L}_{disc},
\end{equation}

\noindent where $\lambda_{pred}$, $\lambda_{id}$,  and $\lambda_{disc}$ are the weight factors. We utilize the total loss $\mathcal{L}$ to train $\mathcal{M}(\cdot;\gamma)$ through the noisy channels under high SNRs to guarantee their basic robustness against channel noises in addition to the semantic editing abilities. In our preliminary experiments, we find that if the training SNRs are low, the editing effects are not satisfying because the extreme channel noises will instead disrupt the learning of semantic editing. Therefore, we train $\mathcal{M}(\cdot;\gamma)$ only under high SNRs and discuss how to fine-tune the JECC codecs under all levels of SNRs in Section \ref{sec:model_fine_tuning}.

\subsection{SNR-aware Channel Coding via Model Fine-tuning}
\label{sec:model_fine_tuning}

Benefiting from the recent successful applications of transfer learning techniques \cite{chen2023vision, chen2024conv}, we propose SNR-aware channel coding based on model fine-tuning. As shown in Figure \ref{fig:overview_ft}, we introduce two lightweight trainable adapters that consist of Fully Connected Layers and do not change the shapes of inputs and outputs to the Joint Editing-Channel Encoder $\mathcal{JECE}(\cdot, \cdot;\gamma)$ and the Joint Semantic-Channel Decoder $\mathcal{JSCD}(\cdot;\theta)$. The complex-valued signal $X$ is sent to the Channel Encoder Adapter $\mathcal{A}_{CE}(\cdot;\phi)$ to obtain the final outcome of channel encoding:
\begin{equation}\label{eq:fine_tune_1}
    X^{\prime} = \mathcal{A}_{CE}(X;\phi).
\end{equation}

\noindent $X^{\prime}$ is then transmitted through the physical channels obeying the similar procedure described in (\ref{eq:channel_transfer}) and the receiver will obtain $Y^{\prime}$. Then, $Y^{\prime}$ is sent to the Channel Decoder Adapter $\mathcal{A}_{CD}(\cdot;\psi)$ to acquire the new input of $\mathcal{JSCD}(\cdot;\theta)$:
\begin{equation}\label{eq:fine_tune_2}
    Y = \mathcal{A}_{CD}(Y^{\prime};\psi).
\end{equation}

In the fine-tuning stage, we randomly select $n_{ft}$ images that strictly have no intersections with the images used in the test time. For each new SNR level that needs to be adapted to, we transmit the fine-tuning images following the procedures defined in Section \ref{system_model} as well as (\ref{eq:fine_tune_1}) and (\ref{eq:fine_tune_2}). We compute the LPIPS loss \cite{zhang2018unreasonable} between the final received image $\hat{S}_{I^{\prime}}$ and the original image $S_I$ to measure the perceptual distortion for the fine-tuning loss $\mathcal{L}_{ft}$. We update the parameters of $\mathcal{A}_{CE}(\cdot;\phi)$ and $\mathcal{A}_{CD}(\cdot;\psi)$ using the gradient descent:
\begin{equation} \label{eq:ft}
\left\{
    \begin{array}{lr}
    \phi^{(t)} = \phi^{(t - 1)} - \eta_{ft} \cdot \nabla_{\phi^{(t - 1)}} \mathcal{L}_{ft} \\
    \psi^{(t)} = \psi^{(t - 1)} - \eta_{ft} \cdot \nabla_{\psi^{(t - 1)}} \mathcal{L}_{ft},
    \end{array}
\right.
\end{equation}

\noindent where $r_{ft}$ is the number of total fine-tuning iterations, $\phi^{(t)}, \psi^{(t)} \; (1 \leq t \leq r_{ft})$ are the results after the $t$-th iteration, and $\eta_{ft}$ is the fine-tuning learning rate. When the fine-tuning is completed, these two adapters can capture the data distribution of current SNR level and enhance the perceptual quality of edited images in the test time, especially under extreme channel conditions. And the rest of the parameters are frozen to avoid forgetting the previously learned priors from Section \ref{sec:jecc} because of the fine-tuning. The algorithm framework of Editable-DeepSC is summarized in Algorithm \ref{alg:training_algorithm}.

\begin{algorithm}[t]
\caption{Editable-DeepSC Training Pipeline}
\label{alg:training_algorithm}
\begin{algorithmic}[1]
\REQUIRE Training dataset $\mathcal{D}$; Pre-trained models $G(\cdot)$, $P(\cdot)$, $F(\cdot)$; Hyper-parameters.
\ENSURE Trained parameters $\gamma, \theta, \phi, \psi$.

\STATE \textbf{Stage 1: Semantic Coding via GAN Inversion}
\STATE \textbf{Goal:} Compress image $S_I$ to latent code $E_I$.
\STATE Encode each $S_I \in \mathcal{D}$ by optimizing $z$ to minimize distortion $J(z)$ between $S_I$ and $G(z)$ as in (\ref{eq:inversion}).
\STATE Obtain semantic code $E_I = z^{(r_{inv})}$.

\STATE \textbf{Stage 2: JECC Training via Iterative Attributes Matching}
\STATE \textbf{Goal:} Learn joint editing-channel coding.
\STATE Encode text $S_T$ to get editing targets $(i_{tar}, d_{tar}, \Delta_{tar})$ via $\mathcal{SE_{T}}(\cdot;\beta)$.
\STATE Iteratively refine $E_I$ using $\mathcal{M}(\cdot;\gamma)$ guided by editing targets as in (\ref{eq:jecc}).
\STATE Train $\gamma, \theta$ by minimizing $\mathcal{L} = \lambda_{pred}\mathcal{L}_{pred} + \lambda_{id}\mathcal{L}_{id} + \lambda_{disc}\mathcal{L}_{disc}$ under high SNRs.

\STATE \textbf{Stage 3: SNR-aware Model Fine-tuning}
\STATE \textbf{Goal:} Adapt to varying channel conditions.
\STATE For each target SNR, fine-tune adapters $\mathcal{A}_{CE}(\cdot;\phi)$ and $\mathcal{A}_{CD}(\cdot;\psi)$ as in (\ref{eq:ft}).
\STATE Update $\phi, \psi$ by minimizing LPIPS loss $\mathcal{L}_{ft}$ on a small image set. Freeze the other parameters.
\end{algorithmic}
\end{algorithm}

\section{Experiments}
\label{experiments}

\subsection{Experiment Setup}

\begin{figure*}[t]
    \centering
    \begin{subfigure}{0.230\linewidth}
        \includegraphics[width=\linewidth]{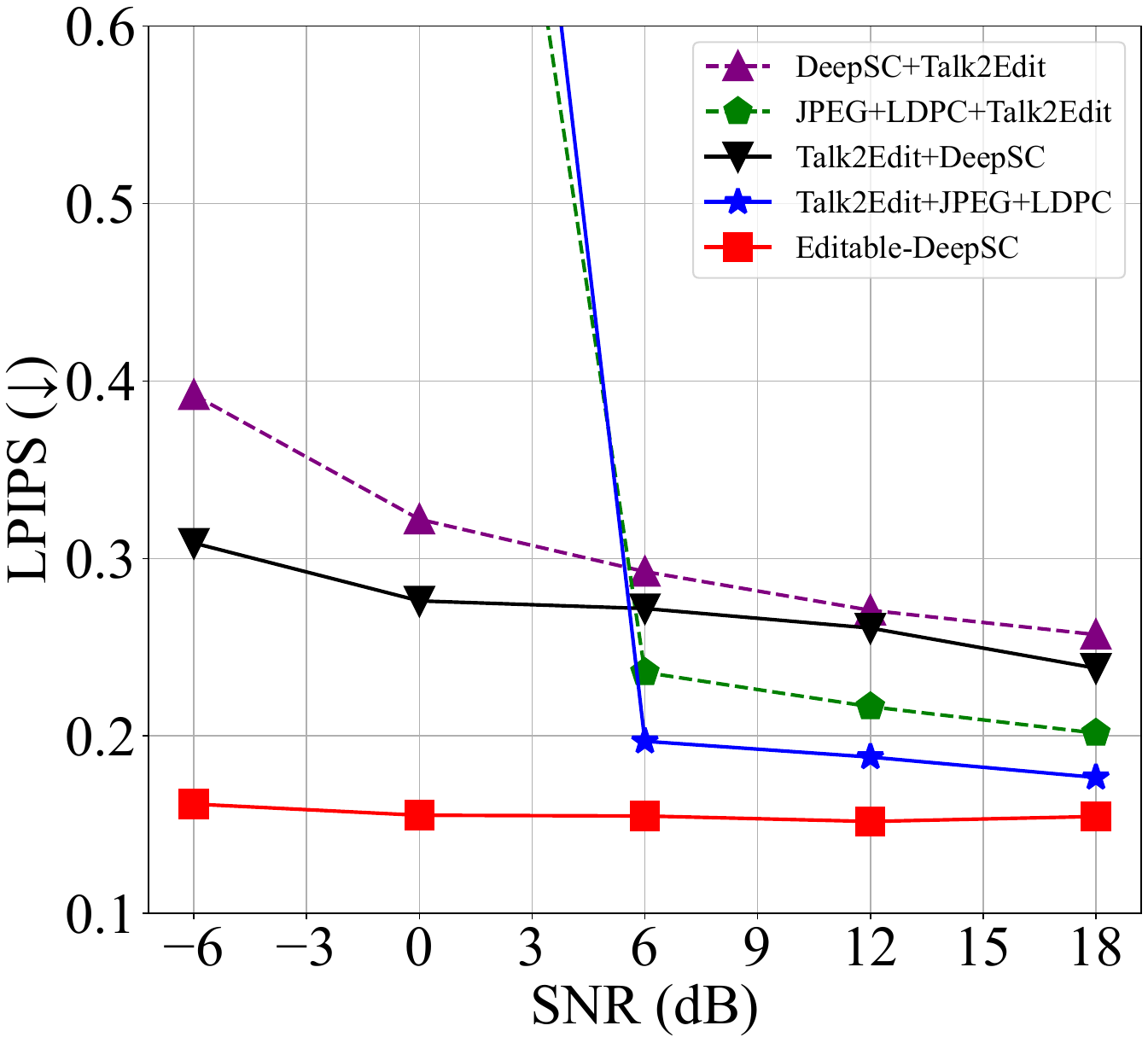}
    \end{subfigure}
    \begin{subfigure}{0.230\linewidth}
        \includegraphics[width=\linewidth]{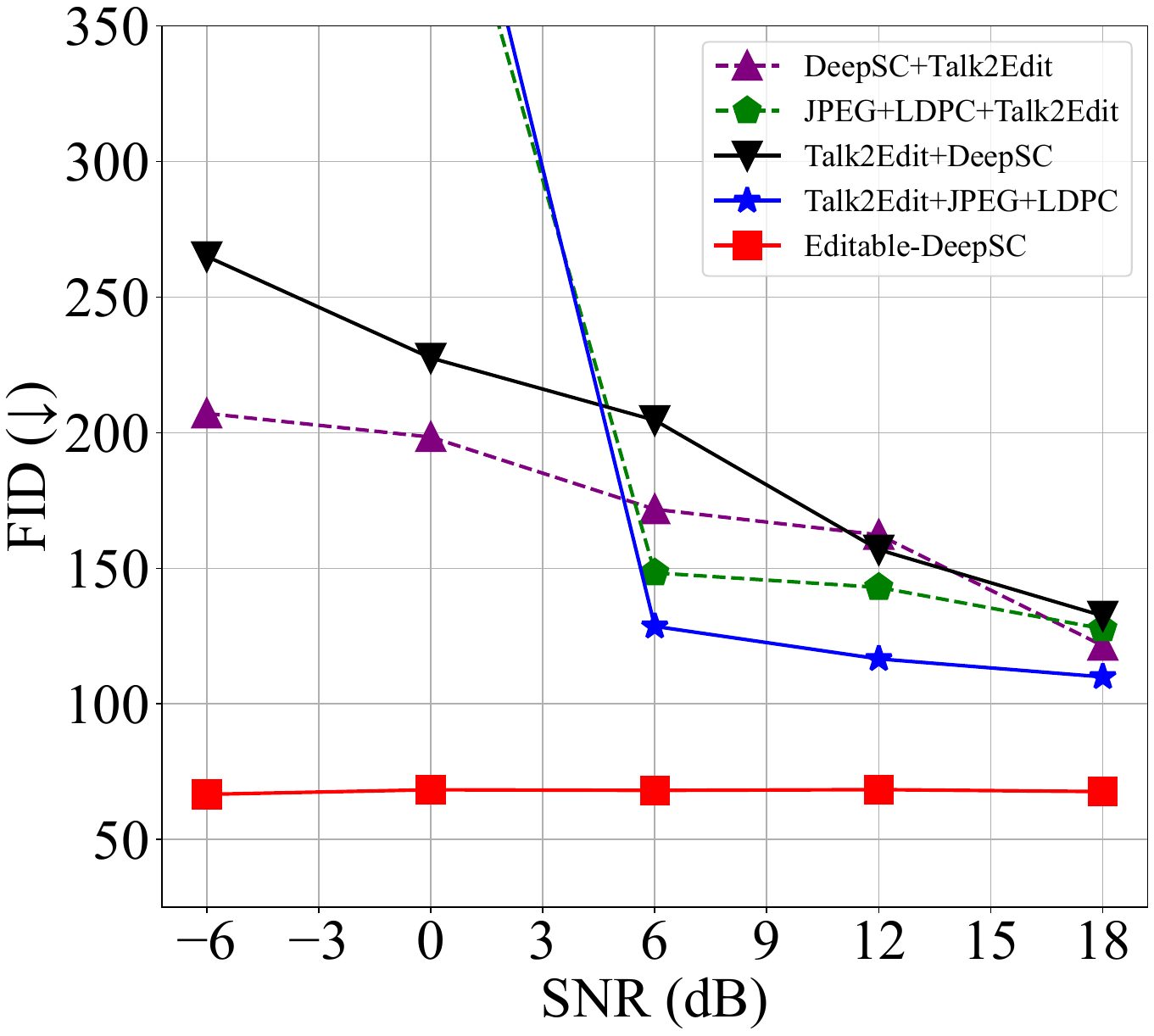}
    \end{subfigure}
    \begin{subfigure}{0.230\linewidth}
        \includegraphics[width=\linewidth]{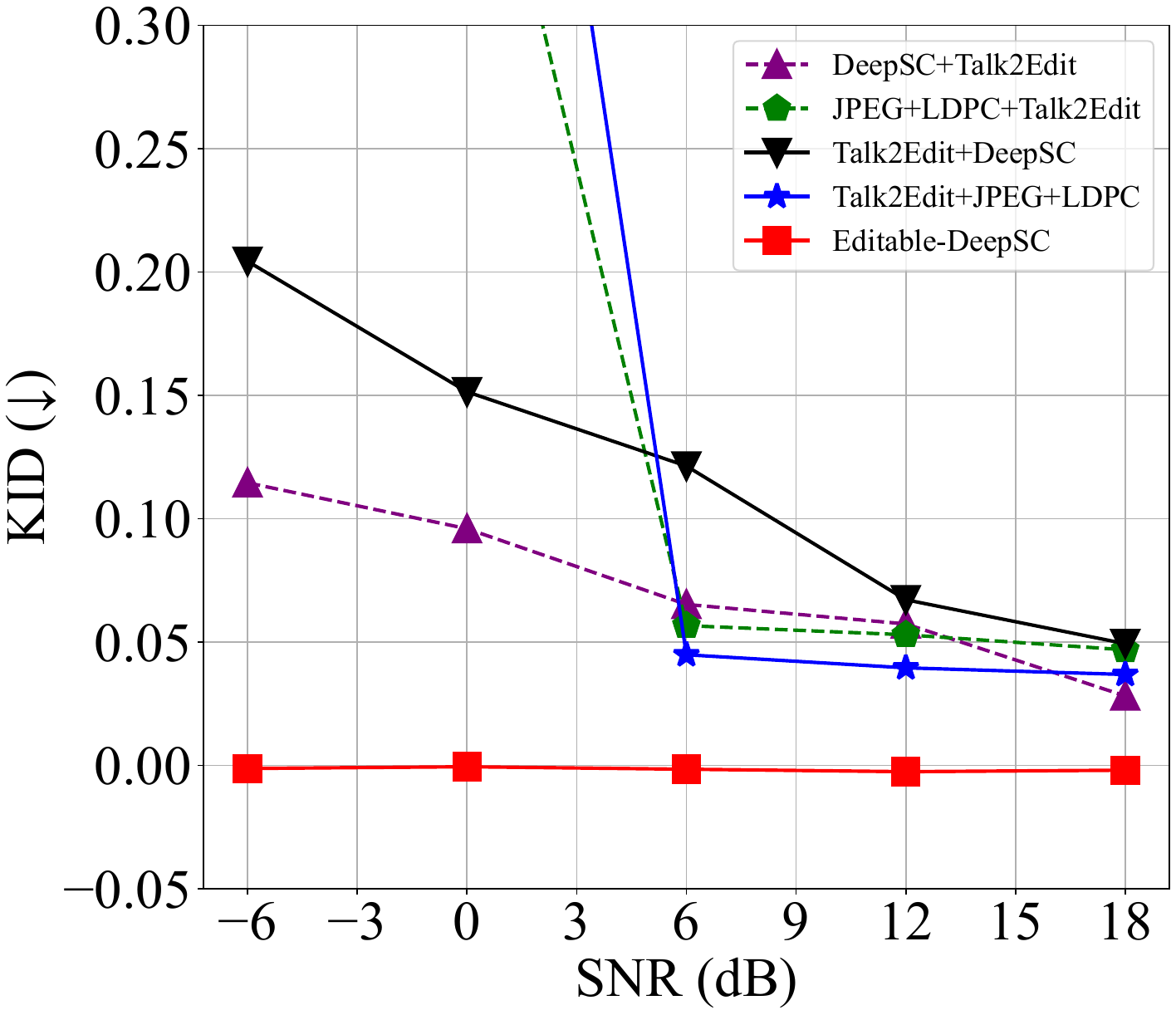}
    \end{subfigure}
    \begin{subfigure}{0.230\linewidth}
        \includegraphics[width=\linewidth]{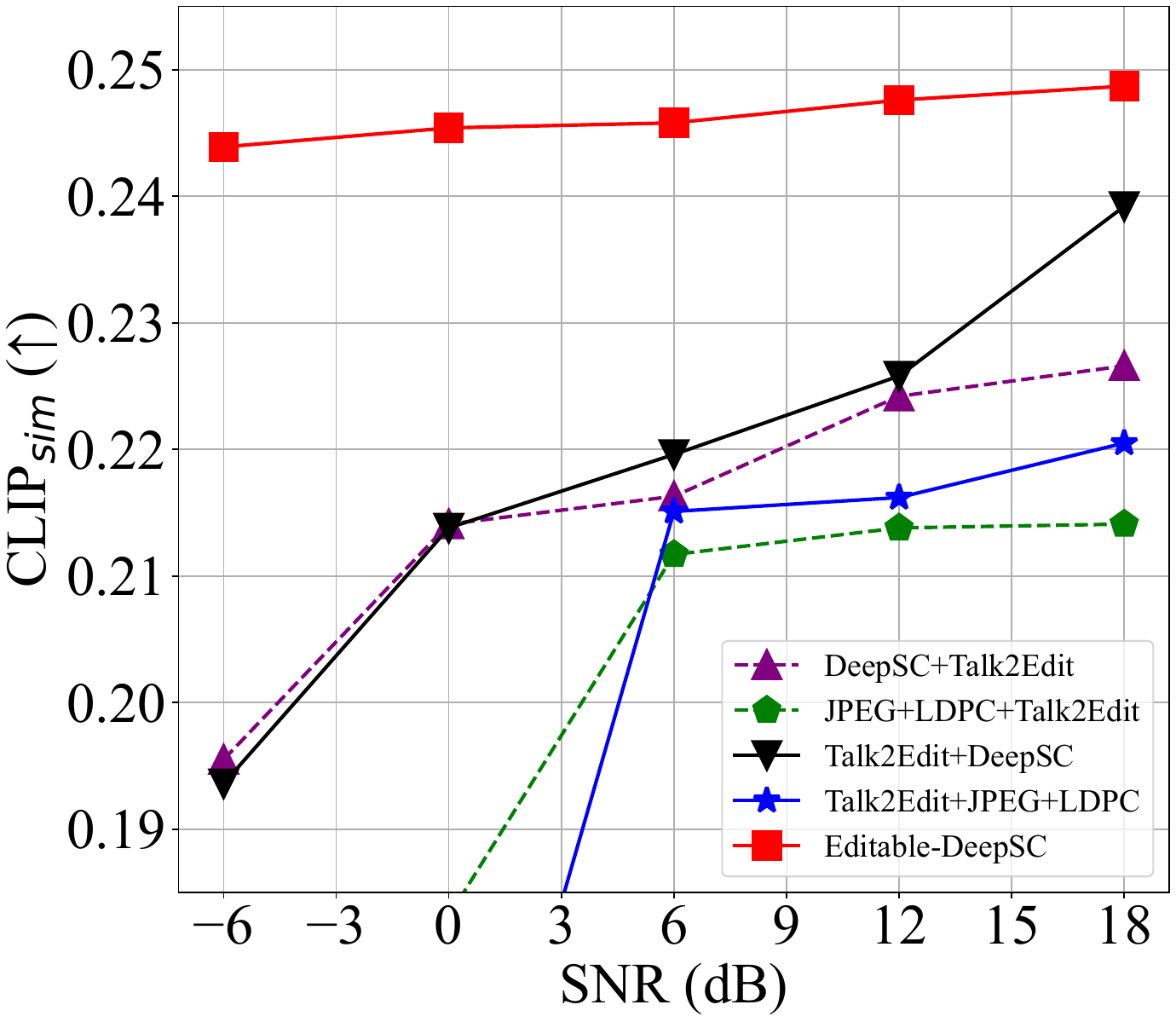}
    \end{subfigure}

    \caption{Quantitative comparison of different methods on the CelebA dataset (resolution $128 \times 128$) for cross-modal language-driven image editing and transmission tasks. Note that $\downarrow$ indicates that the lower the metric, the better the performance, while $\uparrow$ indicates that the higher the metric, the better the performance.}
    \label{fig:main_celeba}
\end{figure*}

\begin{figure*}[t]

        \centering
        \begin{subfigure}{0.92\linewidth}
        \centering
        \begin{minipage}[t]{0.140\linewidth}
        \centering
        \includegraphics[width=2.426cm]{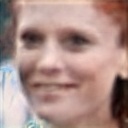}
        \centering
        \end{minipage}
        \begin{minipage}[t]{0.140\linewidth}
        \centering
        \includegraphics[width=2.426cm]{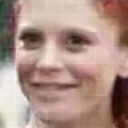}
        \centering
        \end{minipage}
        \begin{minipage}[t]{0.140\linewidth}
        \centering
        \includegraphics[width=2.426cm]{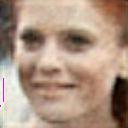}
        \centering
        \end{minipage}
        \begin{minipage}[t]{0.140\linewidth}
        \centering
        \includegraphics[width=2.426cm]{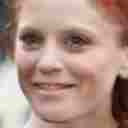}
        \centering
        \end{minipage}
        \begin{minipage}[t]{0.140\linewidth}
        \centering
        \includegraphics[width=2.426cm]{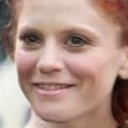}
        \centering
        \end{minipage}
        \begin{minipage}[t]{0.140\linewidth}
        \centering
        \includegraphics[width=2.426cm]{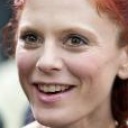}
        \centering
        \end{minipage}
        \end{subfigure}

        \vspace{0.8pt}

        \begin{subfigure}{0.92\linewidth}
        \centering
        \begin{minipage}[t]{0.140\linewidth}
        \centering
        \includegraphics[width=2.426cm]{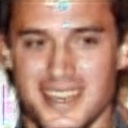}
        \caption*{\footnotesize\textbf{\makecell{DeepSC+Talk2Edit}}}
        \centering
        \end{minipage}
        \begin{minipage}[t]{0.140\linewidth}
        \centering
        \includegraphics[width=2.426cm]{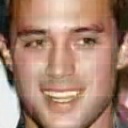}
        \caption*{\footnotesize\textbf{\makecell{JPEG+LDPC\\+Talk2Edit}}}
        \centering
        \end{minipage}
        \begin{minipage}[t]{0.140\linewidth}
        \centering
        \includegraphics[width=2.426cm]{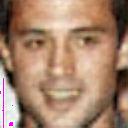}
        \caption*{\footnotesize\textbf{\makecell{Talk2Edit+DeepSC}}}
        \centering
        \end{minipage}
        \begin{minipage}[t]{0.140\linewidth}
        \centering
        \includegraphics[width=2.426cm]{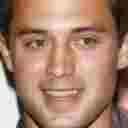}
        \caption*{\footnotesize\textbf{\makecell{Talk2Edit\\+JPEG+LDPC}}}
        \centering
        \end{minipage}
        \begin{minipage}[t]{0.140\linewidth}
        \centering
        \includegraphics[width=2.426cm]{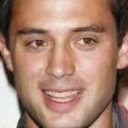}
        \caption*{\footnotesize\textbf{\makecell{Editable-DeepSC}}}
        \centering
        \end{minipage}
        \begin{minipage}[t]{0.140\linewidth}
        \centering
        \includegraphics[width=2.426cm]{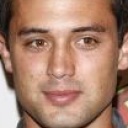}
        \caption*{\footnotesize\textbf{\makecell{Original}}}
        \centering
        \end{minipage}
        \end{subfigure}

    \caption{Qualitative comparison of different methods on the CelebA dataset (resolution $128 \times 128$) for cross-modal language-driven image editing and transmission tasks at the SNR of $6$ dB. The instructive sentences for the $1$st and $2$nd rows are respectively ``I kind of want the smile to be \underline{less} obvious" and ``Smile \underline{more}".}
    \label{fig:visualization_main}
\end{figure*}

\textbf{Simulation Datasets.} As for the textual instruction $S_T$, we use the editing requests from the CelebA-Dialog dataset \cite{jiang2021talk} for all the experiments. As for the original image $S_I$: (1) In the main experiments, we consider relatively simple scenarios and choose from the CelebA dataset \cite{liu2015deep}, whose images are cropped to the resolution of $128 \times 128$ and have similar distribution with the training set of StyleGAN \cite{karras2019style}; (2) In the more realistic high-resolution experiments, we choose from the CelebA-HQ dataset \cite{karras2018progressive} with the resolution of $1024 \times 1024$; (3) In the more realistic out-of-distribution (OOD) experiments, we choose from the MetFaces dataset \cite{karras2020training} with the resolution of $128 \times 128$, whose facial images are from artworks and have an inherent distribution bias with the natural facial images of StyleGAN training set. As for the fine-tuning images, we use images from the CelebA dataset \cite{liu2015deep} for the resolution of $128 \times 128$, and use images from the CelebA-HQ dataset \cite{karras2018progressive} for the resolution of $1024 \times 1024$. We strictly ensure that the fine-tuning images are few in number ($n_{ft} < 20$) and have no intersections with the tested images.

\textbf{Evaluation Baselines.} To the best of our knowledge, there exists no other semantic communication approach tailored for cross-modal facial editing tasks by now. Therefore, we compare Editable-DeepSC with the classical traditional communication methods JPEG \cite{wallace1992jpeg} and LDPC \cite{gallager1962low}, the novel single-modal image semantic communication method DeepSC \cite{zhang2022deep}, and the SOTA text-driven image editing method Talk2Edit \cite{jiang2023talk} in the computer vision community. We rearrange these methods and consider the following baselines that separately handle communications and editings:

\begin{itemize}
    \item \textbf{DeepSC+Talk2Edit.} DeepSC \cite{zhang2022deep} is utilized to send $S_I$ from the transmitter to the receiver. And the transmission of $S_T$ is assumed to be error-free, which means that the received sentences are identical to the original ones to enhance the capabilities of this baseline for better persuasiveness. Then, Talk2Edit \cite{jiang2023talk} is utilized to perform the text-driven image editing at the receiver side. This consists with the scheme \emph{$M_1$} mentioned in Section \ref{sec:theoretical_analysis}.
    \item \textbf{JPEG+LDPC+Talk2Edit.} JPEG \cite{wallace1992jpeg} is utilized for image source coding and LDPC \cite{gallager1962low} is utilized for image channel coding to send $S_I$ from the transmitter to the receiver. And the transmission of $S_T$ is also assumed to be error-free. Then, Talk2Edit \cite{jiang2023talk} is utilized to perform the text-driven image editing at the receiver side. This consists with the scheme \emph{$M_1$} mentioned in Section \ref{sec:theoretical_analysis}.
    \item \textbf{Talk2Edit+DeepSC.} Talk2Edit \cite{jiang2023talk} is utilized to perform the text-driven image editing at the transmitter side. Then, DeepSC \cite{zhang2022deep} is utilized to send the edited images from the transmitter to the receiver. This consists with the scheme \emph{$M_2$} mentioned in Section \ref{sec:theoretical_analysis}.
    \item \textbf{Talk2Edit+JPEG+LDPC.} Talk2Edit \cite{jiang2023talk} is utilized to perform the text-driven image editing at the transmitter side. Then, JPEG \cite{wallace1992jpeg} is utilized for image source coding and LDPC \cite{gallager1962low} is utilized for image channel coding to send the edited images from the transmitter to the receiver. This consists with the scheme \emph{$M_2$} mentioned in Section \ref{sec:theoretical_analysis}.
\end{itemize}

\textbf{Quantitative Metrics.} To quantitatively measure the editing effects of different methods, we adopt LPIPS \cite{zhang2018unreasonable}, FID \cite{heusel2017gans}, KID \cite{binkowski2018demystifying}, and CLIP$_{sim}$ \cite{radford2021learning}. LPIPS utilizes pre-trained convolutional networks to extract the features of original images and edited images to calculate the perceptual differences. FID and KID computes the distribution differences between the original images and edited images. CLIP$_{sim}$ first utilizes the pre-trained CLIP models \cite{radford2021learning} to extract both the image and text features, and then computes the cosine similarity between the image features of edited results and the text features of target prompts to evaluate the semantic alignment with the editing instructions. Smaller LPIPS, FID, and KID scores indicate better visual fidelity, while higher CLIP$_{sim}$ scores indicate better editing effects. To quantitatively measure the communication efficiency, we adopt the Channel Bandwidth Ratio (CBR) defined in (\ref{eq:CBR}). Smaller CBR indicates better compression and communication efficiency.

\textbf{Implementation Details.} To simulate the varying communication circumstances, we set the SNR levels of $-6$ dB, $0$ dB, $6$ dB, $12$ dB, $18$ dB for all the methods. For Editable-DeepSC, $\lambda_{inv}$, $\eta_{inv}$, $r_{inv}$, $\eta_{ft}$, $r_{ft}$, and $n_{ft}$ are set as $1.0$, $0.10$, $110$, $1\times10^{-4}$, $6$, and $10$, respectively. For DeepSC \cite{zhang2022deep}, the learning rate is set as $1\times10^{-3}$ and the batch size is set as $4$. For JPEG \cite{wallace1992jpeg} and LDPC \cite{gallager1962low}, the target bpp under the resolution of $128\times128$ is set as $0.7$, the target bpp under the resolution of $1024\times1024$ is set as $0.2$, and the Quadrature Amplitude Modulation (QAM) order is set as $4$. For Talk2Edit \cite{jiang2023talk}, we adopt its official code implementation. To ensure a fair comparison, the learning-based transmission approach, DeepSC \cite{zhang2022deep}, is first fully trained under each considered SNR level before testing. All the experiments are conducted on NVIDIA RTX A6000 GPUs.

\subsection{Main Results}
\label{sec:main_results}

\begin{figure*}[h]
    \centering
    \begin{subfigure}{0.230\linewidth}
        \includegraphics[width=\linewidth]{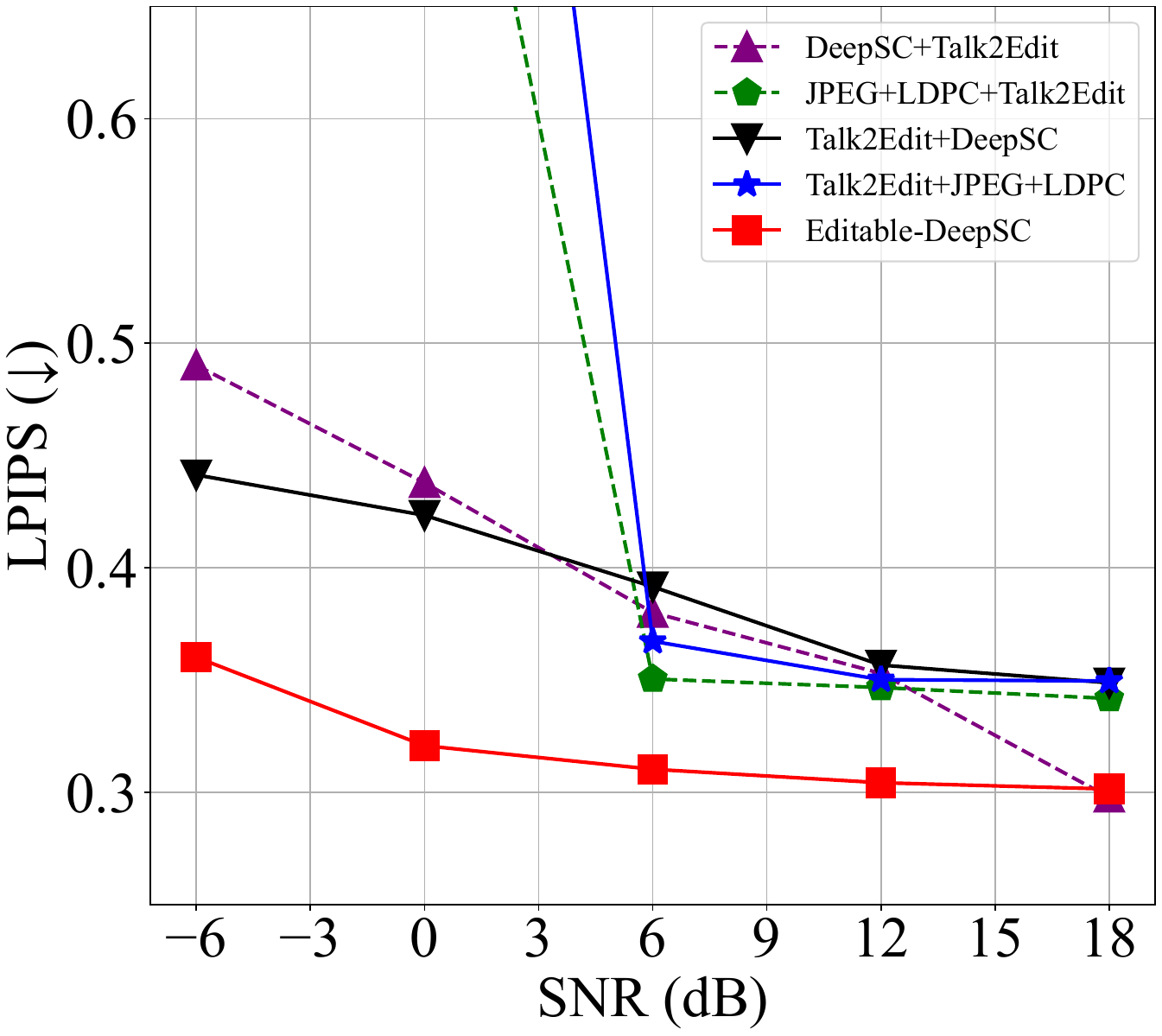}
    \end{subfigure}
    \begin{subfigure}{0.230\linewidth}
        \includegraphics[width=\linewidth]{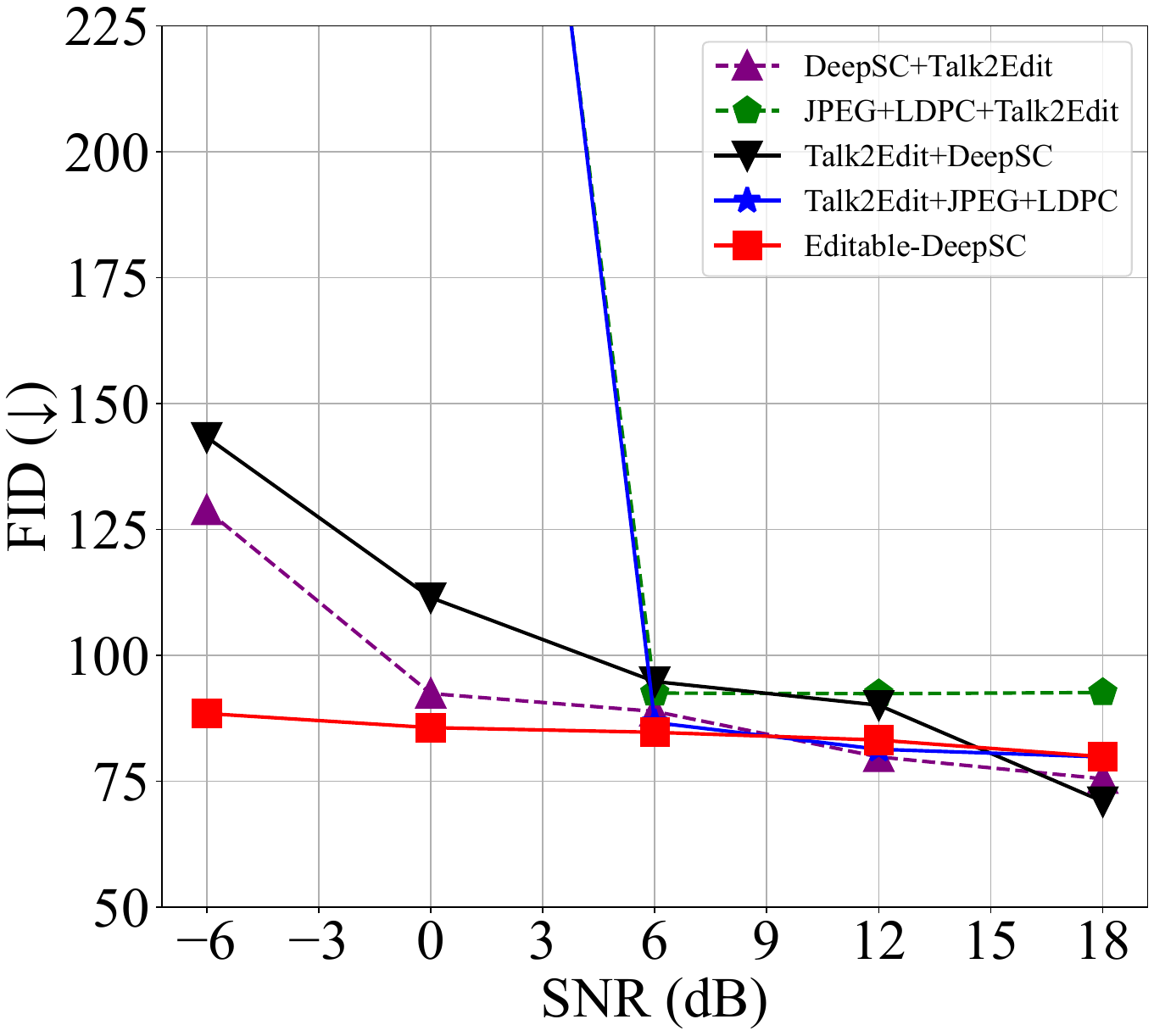}
    \end{subfigure}
    \begin{subfigure}{0.230\linewidth}
        \includegraphics[width=\linewidth]{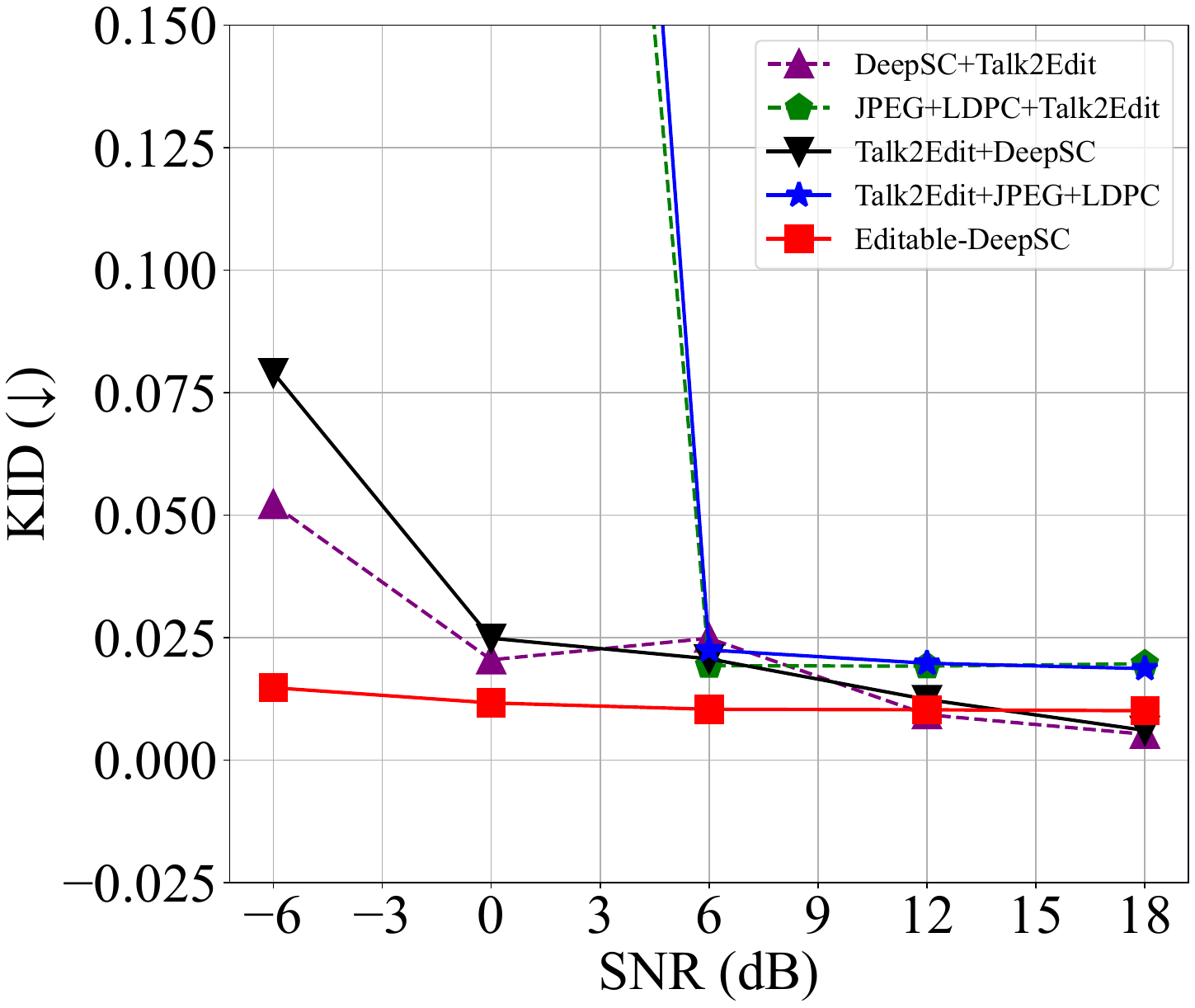}
    \end{subfigure}
    \begin{subfigure}{0.230\linewidth}
        \includegraphics[width=\linewidth]{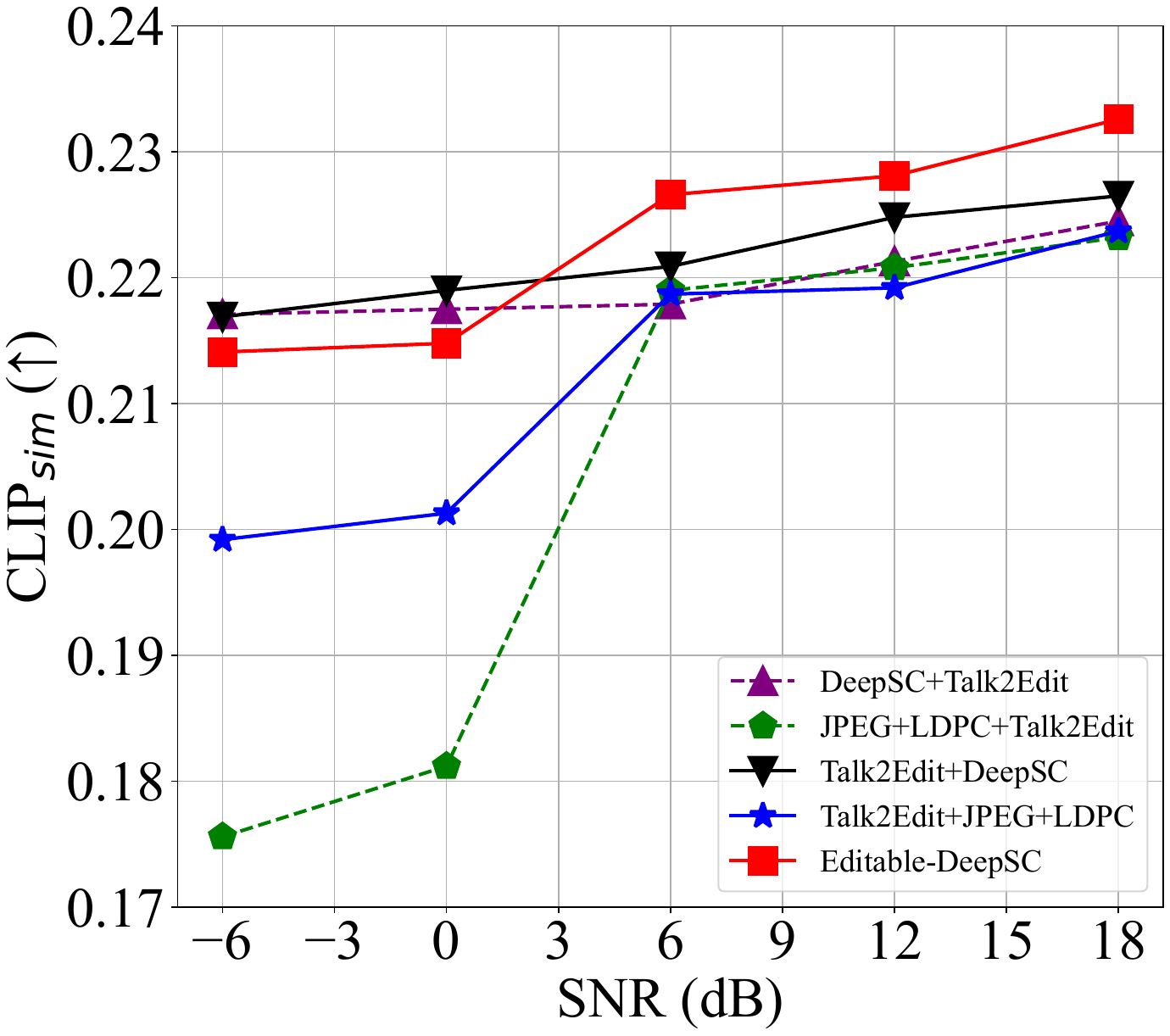}
    \end{subfigure}

    \caption{Quantitative comparison of different methods on the CelebA-HQ dataset (resolution $1024 \times 1024$) for cross-modal language-driven image editing and transmission tasks. Note that $\downarrow$ indicates that the lower the metric, the better the performance, while $\uparrow$ indicates that the higher the metric, the better the performance.}
    \label{fig:more_celeba_hq}
\end{figure*}

\begin{figure*}[h]

        \centering
        \begin{subfigure}{0.92\linewidth}
        \centering
        \begin{minipage}[t]{0.140\linewidth}
        \centering
        \includegraphics[width=2.426cm]{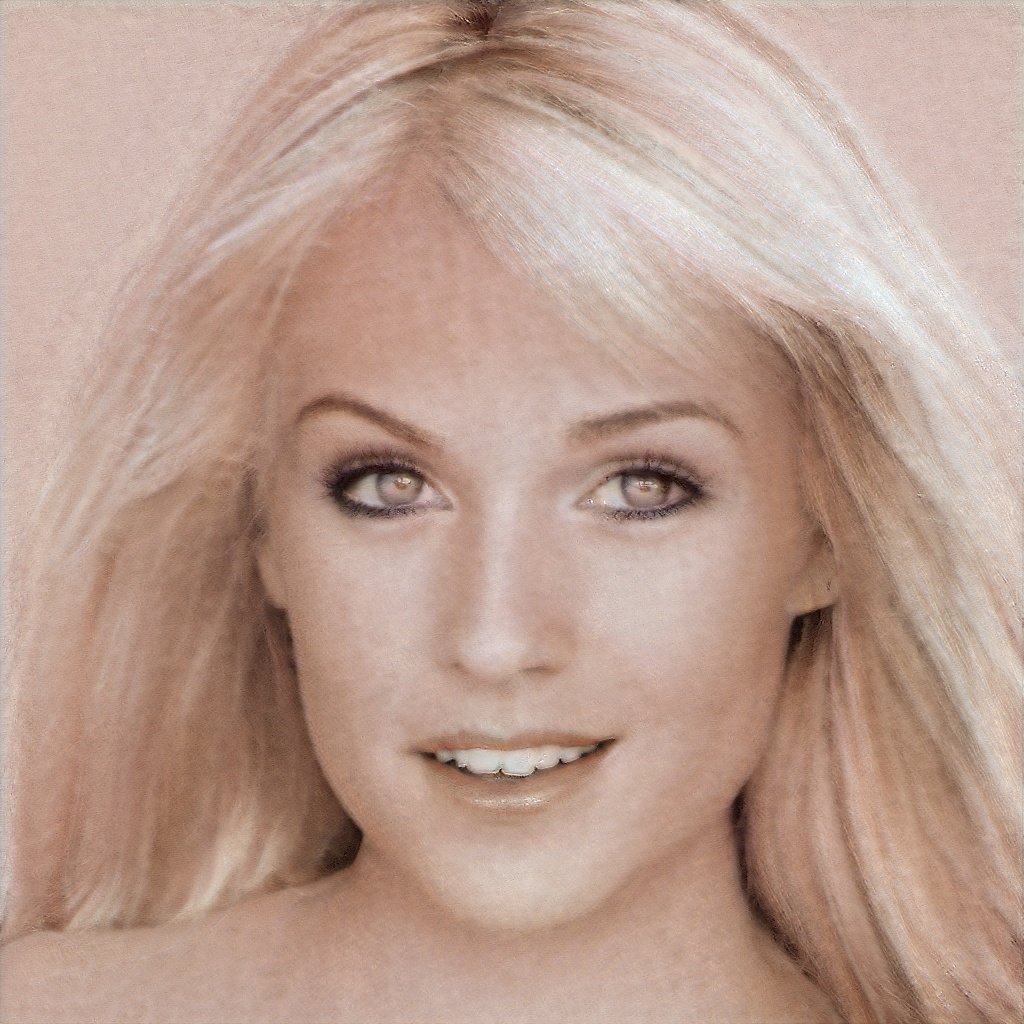}
        \centering
        \end{minipage}
        \begin{minipage}[t]{0.140\linewidth}
        \centering
        \includegraphics[width=2.426cm]{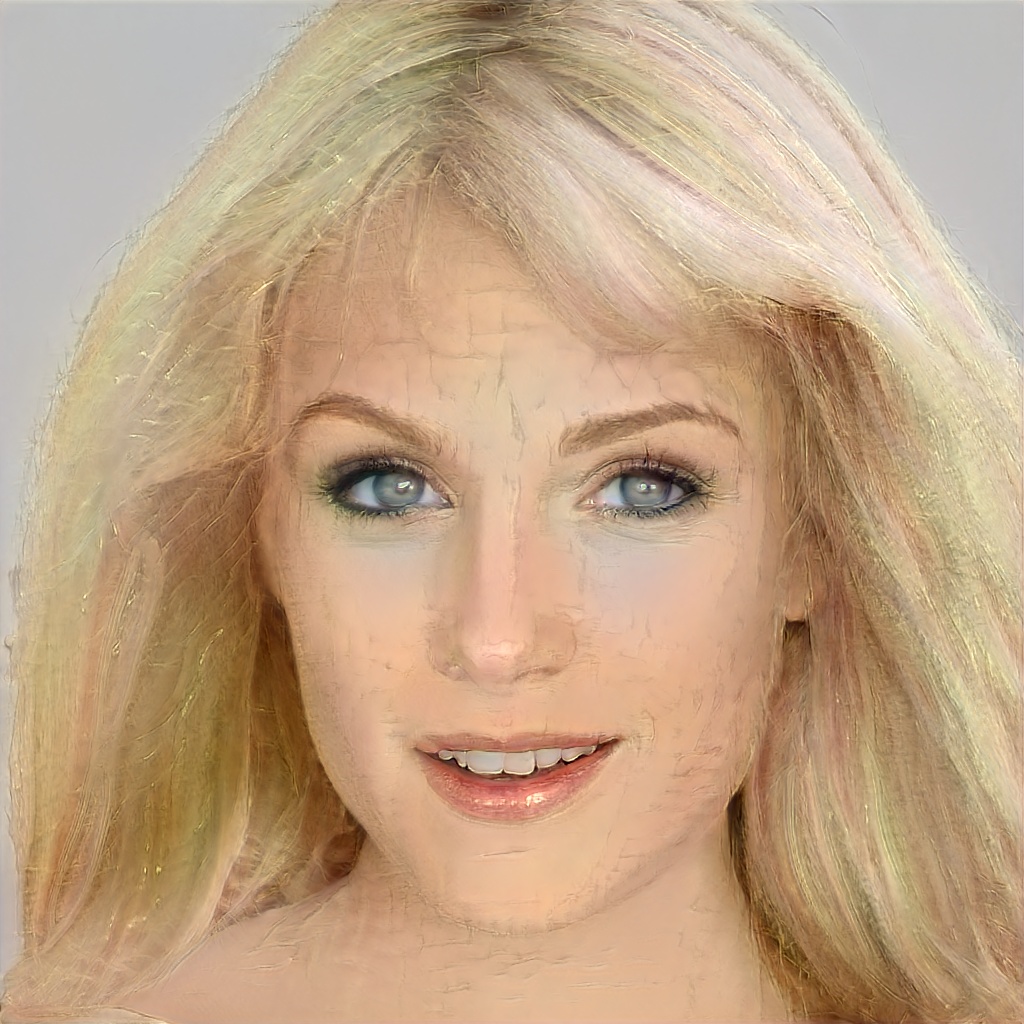}
        \centering
        \end{minipage}
        \begin{minipage}[t]{0.140\linewidth}
        \centering
        \includegraphics[width=2.426cm]{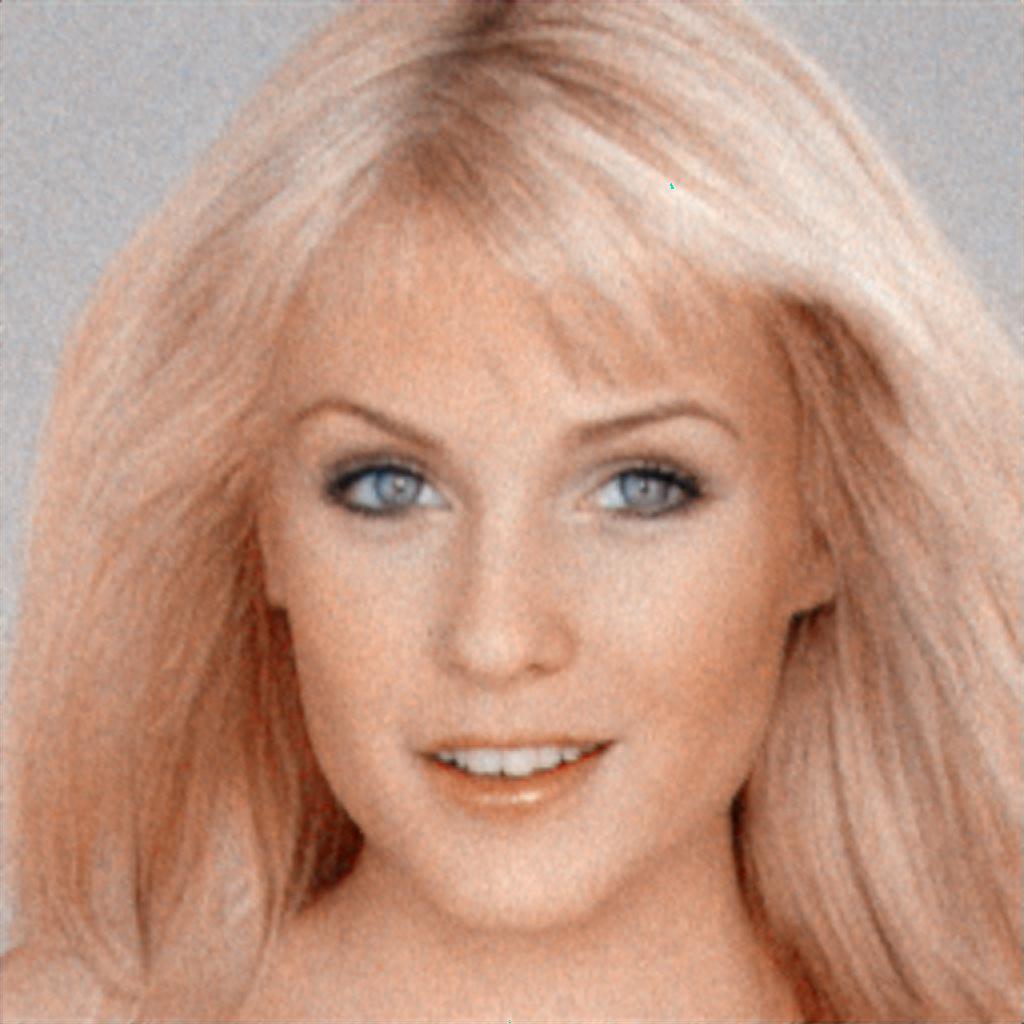}
        \centering
        \end{minipage}
        \begin{minipage}[t]{0.140\linewidth}
        \centering
        \includegraphics[width=2.426cm]{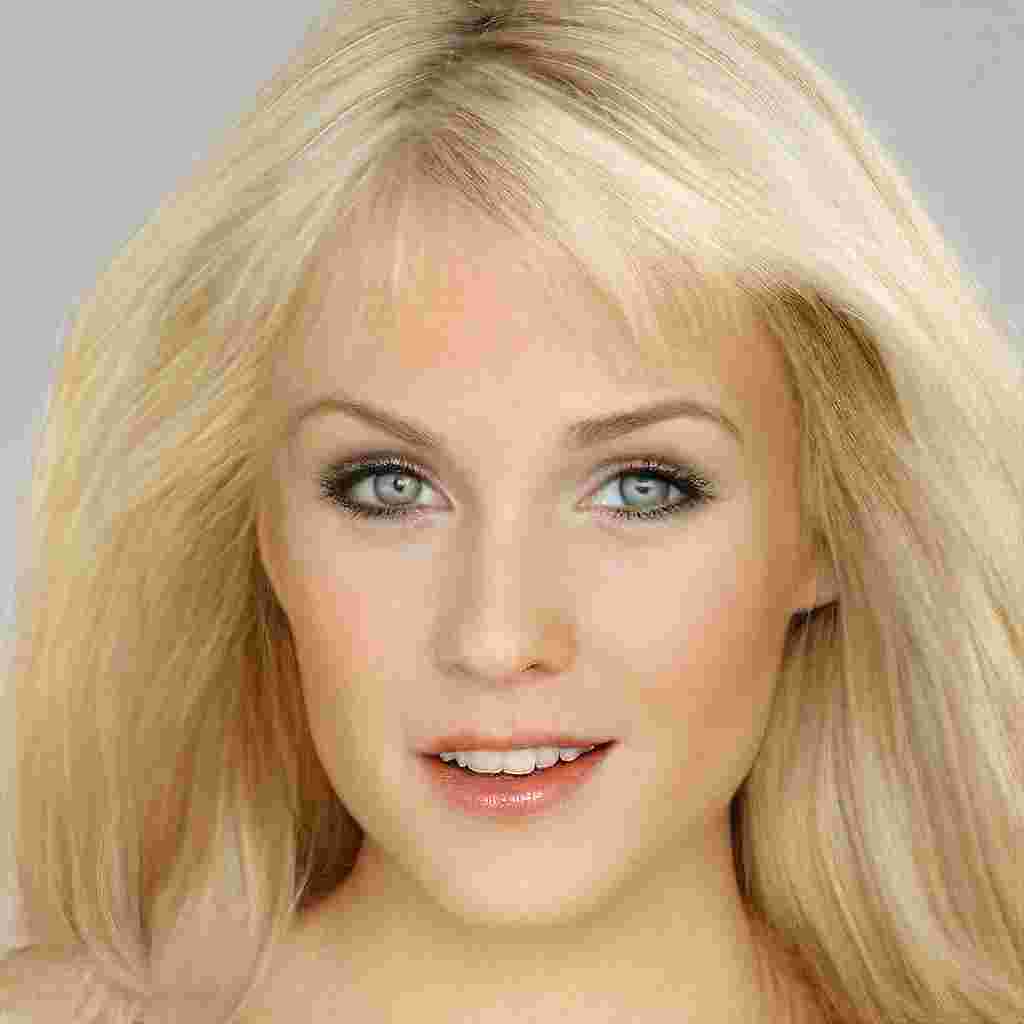}
        \centering
        \end{minipage}
        \begin{minipage}[t]{0.140\linewidth}
        \centering
        \includegraphics[width=2.426cm]{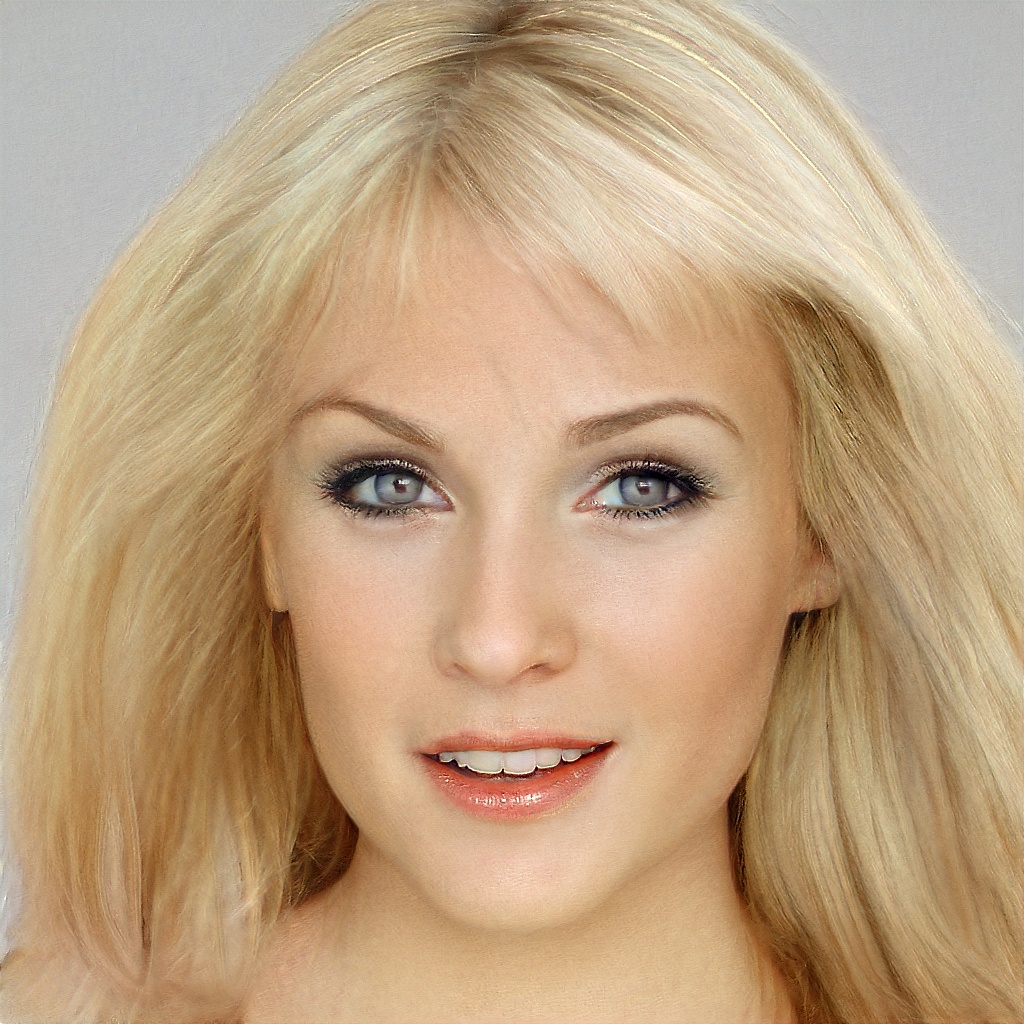}
        \centering
        \end{minipage}
        \begin{minipage}[t]{0.140\linewidth}
        \centering
        \includegraphics[width=2.426cm]{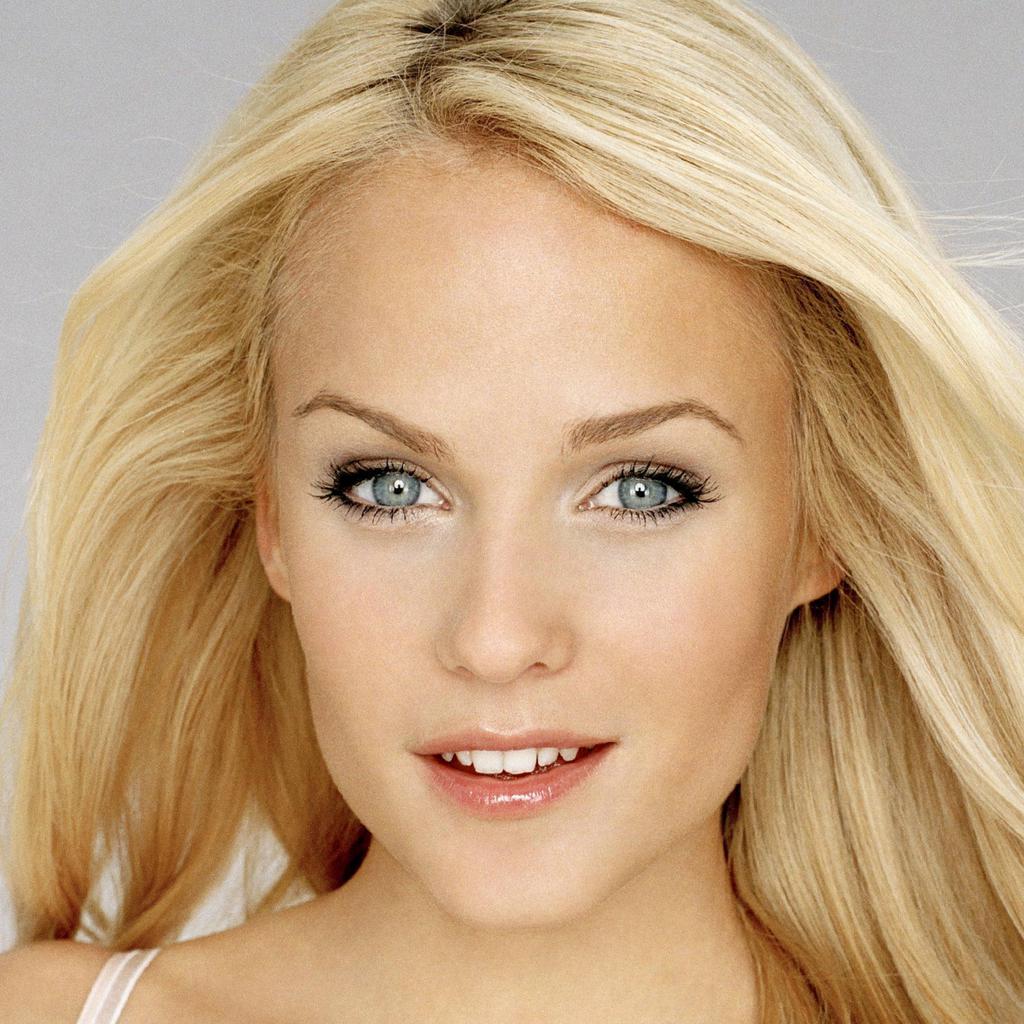}
        \centering
        \end{minipage}
        \end{subfigure}

        \vspace{0.8pt}

        \begin{subfigure}{0.92\linewidth}
        \centering
        \begin{minipage}[t]{0.140\linewidth}
        \centering
        \includegraphics[width=2.426cm]{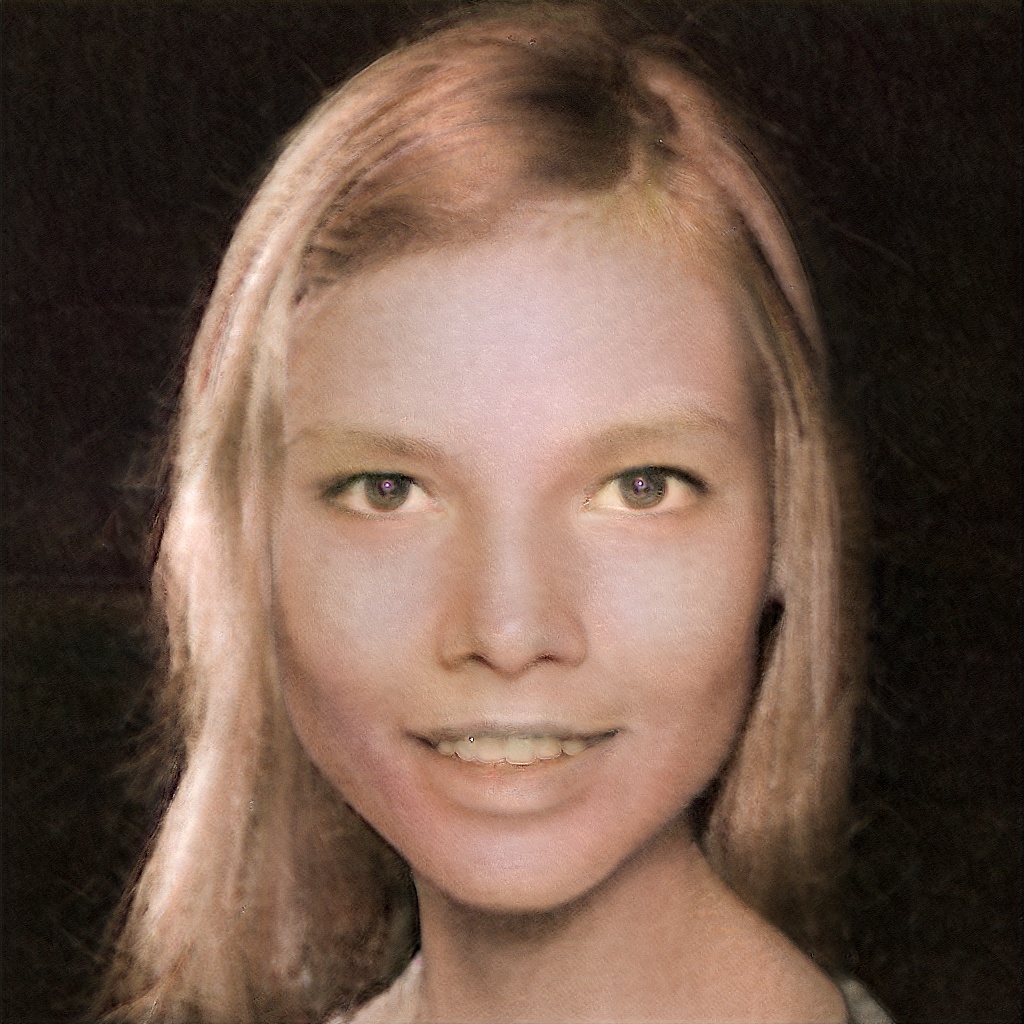}
        \caption*{\footnotesize\textbf{\makecell{DeepSC+Talk2Edit}}}
        \centering
        \end{minipage}
        \begin{minipage}[t]{0.140\linewidth}
        \centering
        \includegraphics[width=2.426cm]{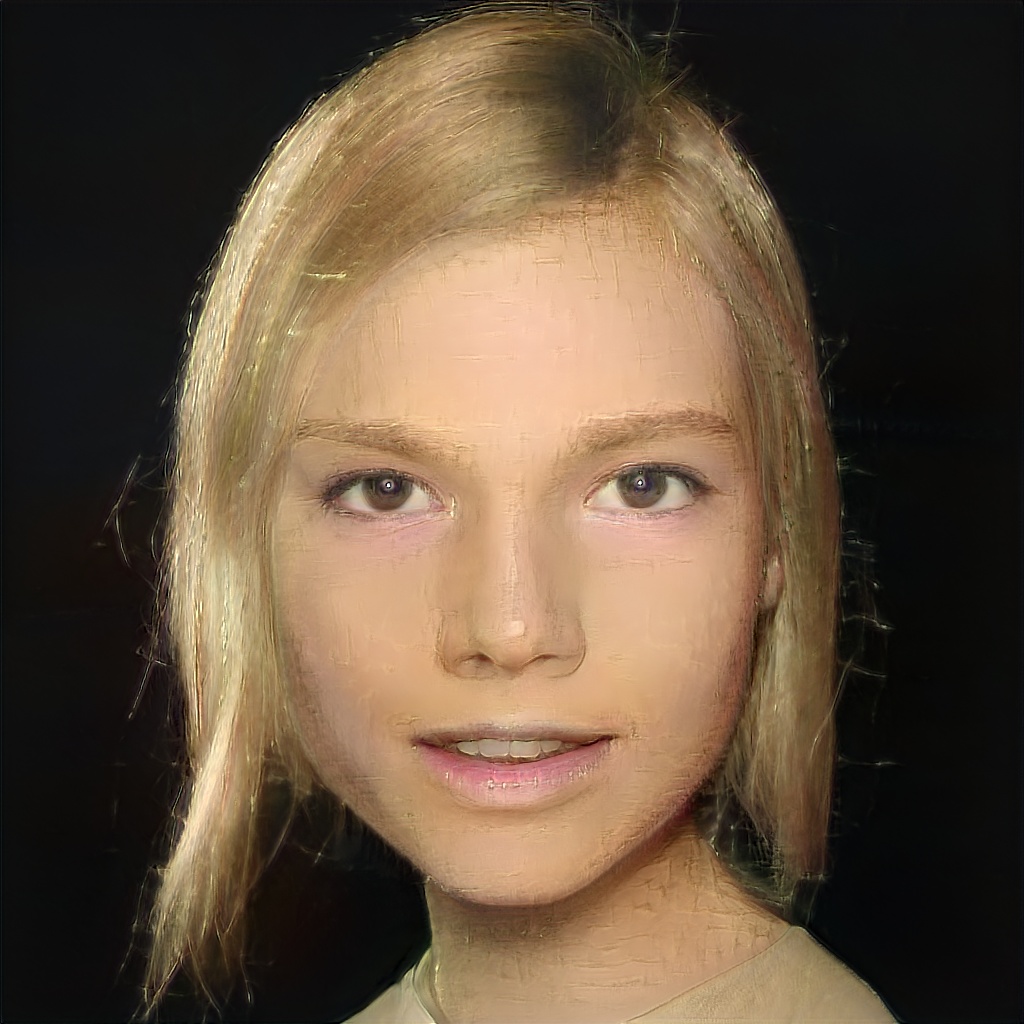}
        \caption*{\footnotesize\textbf{\makecell{JPEG+LDPC\\+Talk2Edit}}}
        \centering
        \end{minipage}
        \begin{minipage}[t]{0.140\linewidth}
        \centering
        \includegraphics[width=2.426cm]{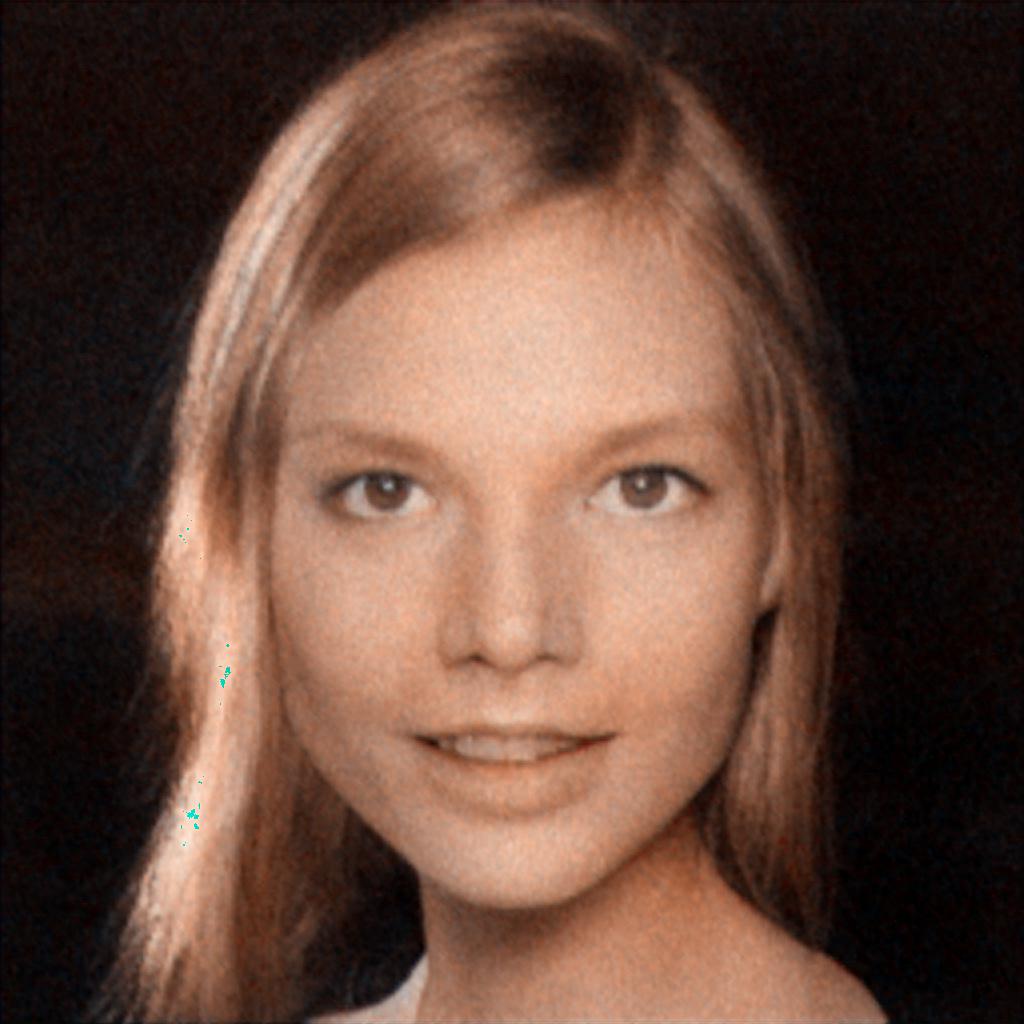}
        \caption*{\footnotesize\textbf{\makecell{Talk2Edit+DeepSC}}}
        \centering
        \end{minipage}
        \begin{minipage}[t]{0.140\linewidth}
        \centering
        \includegraphics[width=2.426cm]{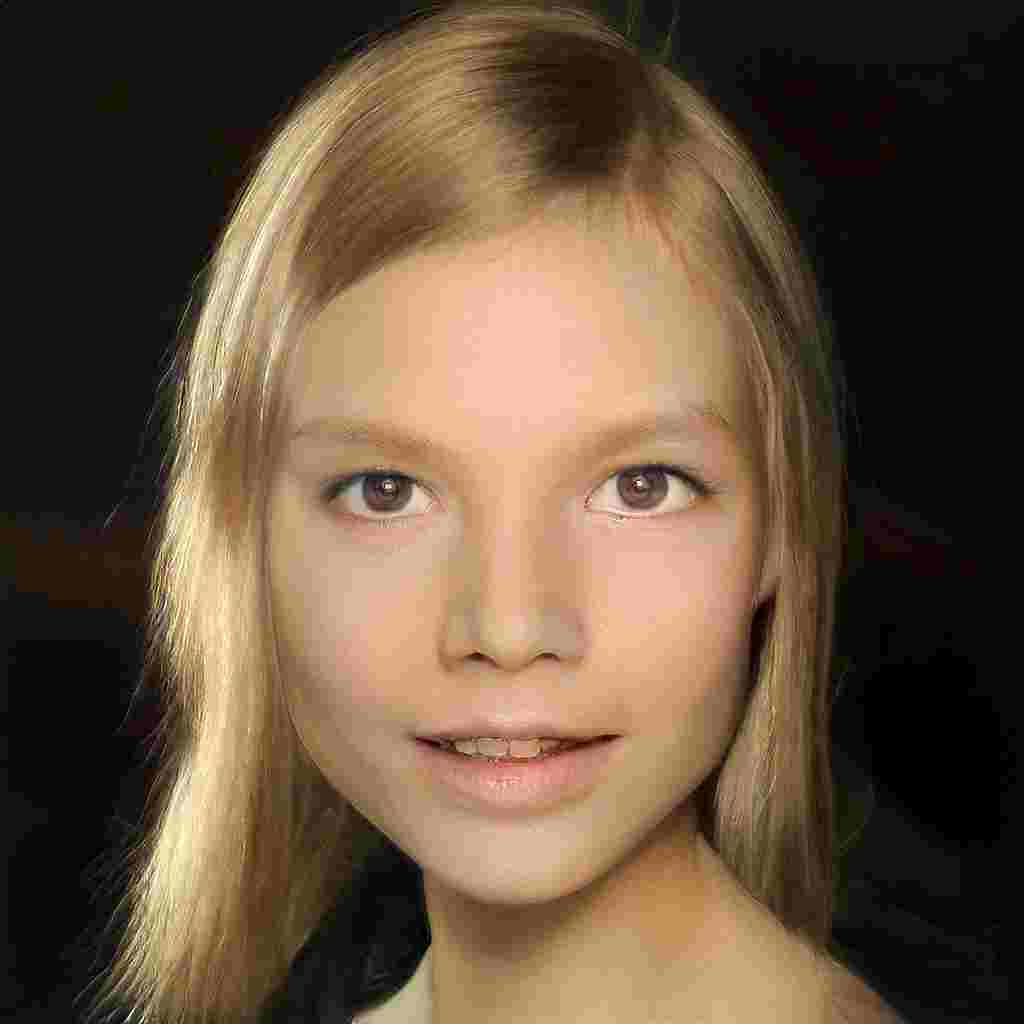}
        \caption*{\footnotesize\textbf{\makecell{Talk2Edit\\+JPEG+LDPC}}}
        \centering
        \end{minipage}
        \begin{minipage}[t]{0.140\linewidth}
        \centering
        \includegraphics[width=2.426cm]{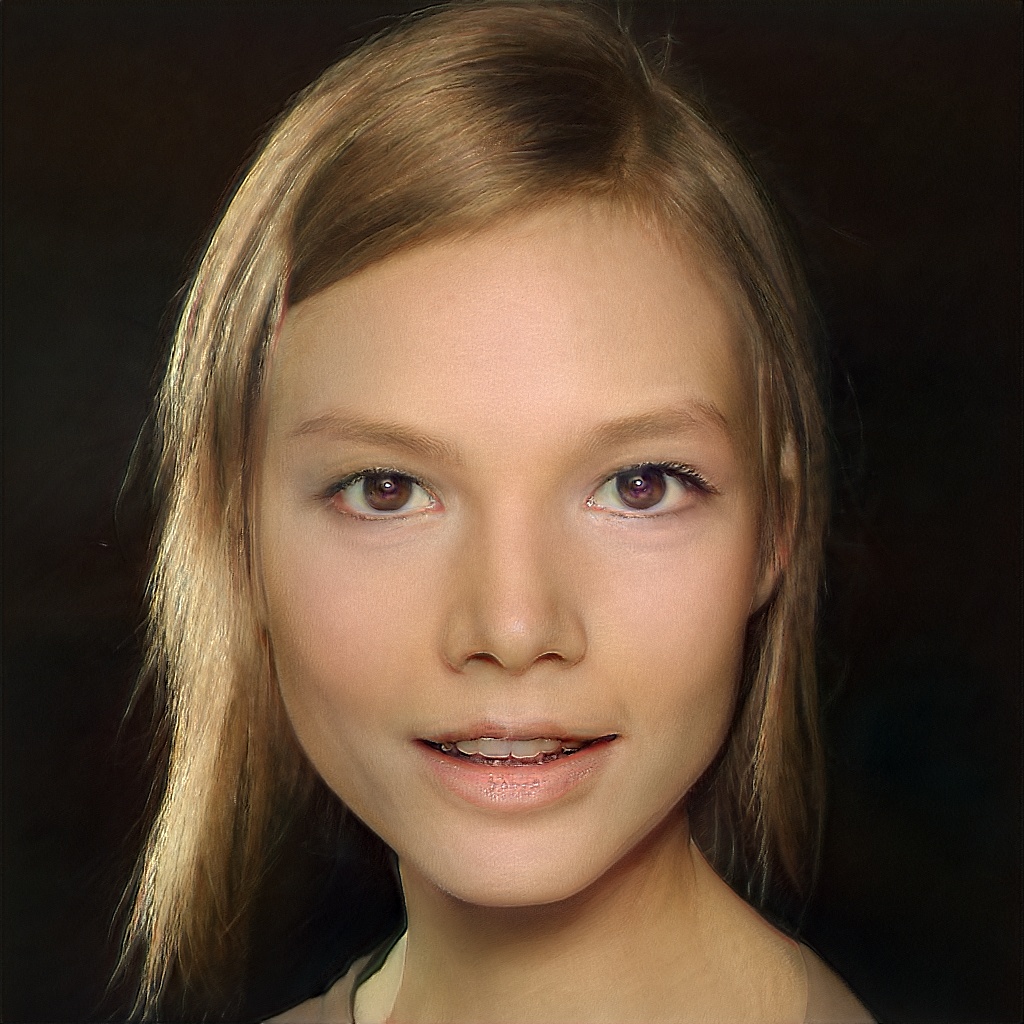}
        \caption*{\footnotesize\textbf{\makecell{Editable-DeepSC}}}
        \centering
        \end{minipage}
        \begin{minipage}[t]{0.140\linewidth}
        \centering
        \includegraphics[width=2.426cm]{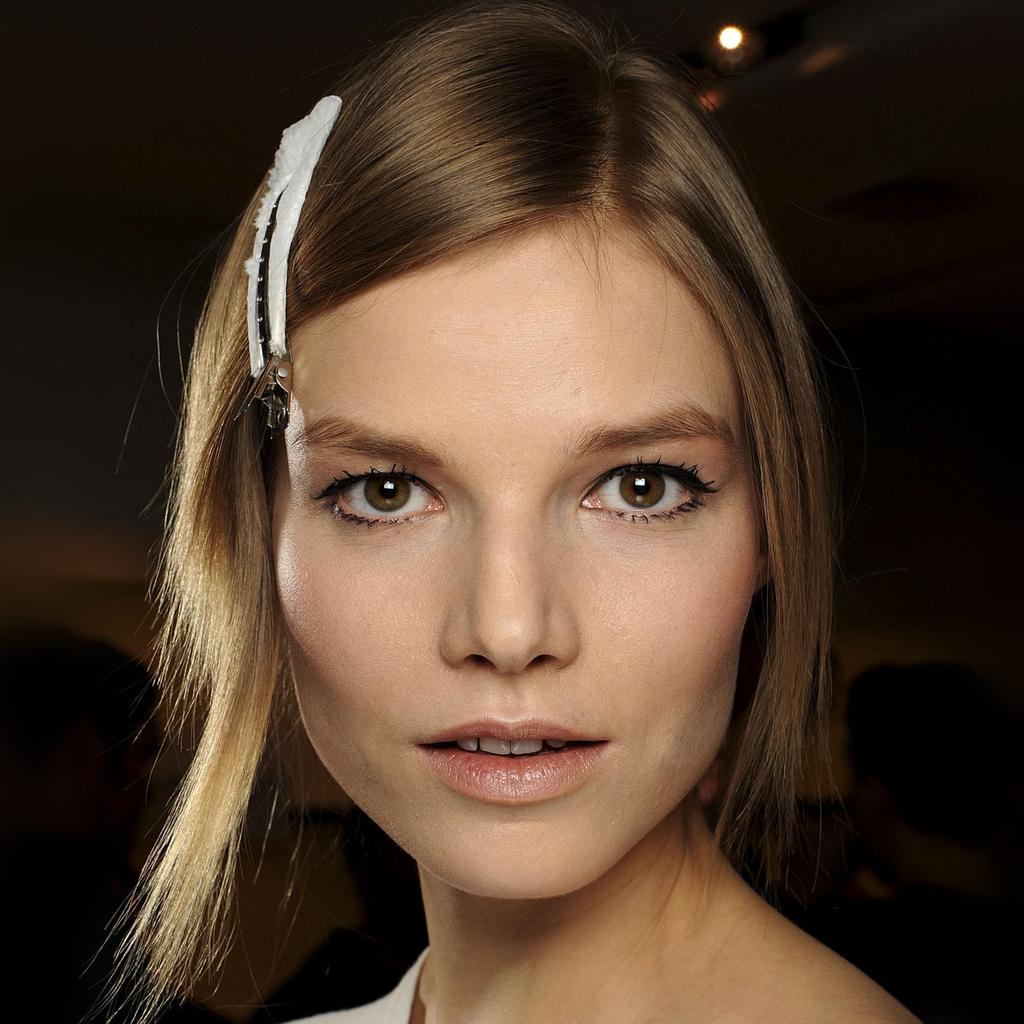}
        \caption*{\footnotesize\textbf{\makecell{Original}}}
        \centering
        \end{minipage}
        \end{subfigure}

    \caption{Qualitative comparison of different methods on the CelebA-HQ dataset (resolution $1024 \times 1024$) for cross-modal language-driven image editing and transmission tasks at the SNR of $6$ dB. The instructive sentences for the $1$st and $2$nd rows are respectively ``What about trying bangs that leaves \underline{half of} the forehead visible" and ``Make the face \underline{slightly younger}".}
    \label{fig:visualization_more_celeba_hq}
\end{figure*}

We compare Editable-DeepSC with the baselines on the CelebA dataset. From Figure \ref{fig:main_celeba}, we find that Editable-DeepSC realizes the best results in terms of LPIPS, FID, KID, and CLIP$_{sim}$ with significant gains. For instance, when the SNR level is $-6$ dB, the LPIPS, FID, KID, and CLIP$_{sim}$ scores of Editable-DeepSC are improved by $47.65\%$, $67.82\%$, $101.05\%$, and $24.76\%$ than the second best method, respectively. These results indicate that Editable-DeepSC achieves excellent editing effects with high fidelity and quality, outperforming the baselines that separately handle communications and editings. We also observe that the performance of JPEG+LDPC+Talk2Edit and Talk2Edit+JPEG+LDPC will decrease rapidly when the SNR level is below $6$ dB, which is consistent with the fact that the traditional communications suffer greatly from the \emph{cliff effect} \cite{bourtsoulatze2019deep}.

\begin{table}
    \caption{Compression effectiveness of different methods with the resolution of $128 \times 128$, measured by the Channel Bandwidth Ratio (CBR) defined in (\ref{eq:CBR}).}
    \label{tab:main_cbr}
    \centering
    \adjustbox{max width=0.5\textwidth}{\begin{tabular}{|c|c|c|}
    \hline
    Method & Resolution & CBR ($\downarrow$) \\
    \hline
    DeepSC+Talk2Edit & $128 \times 128$ & 0.0833 \\
    JPEG+LDPC+Talk2Edit & $128 \times 128$ & 0.0389 \\
    Talk2Edit+DeepSC & $128 \times 128$ & 0.0833 \\
    Talk2Edit+JPEG+LDPC & $128 \times 128$ & 0.0389 \\
    \gc{Editable-DeepSC} & \gc$128 \times 128$ & \gc\textbf{0.0104} \\
    \hline
    \end{tabular}}
\end{table}

Figure \ref{fig:visualization_main} illustrates the qualitative comparison of different methods at the noise level of $6$ dB. We notice that Editable-DeepSC indeed achieves the best visualization results. Although the other methods also manage to edit the images, their effects are not as vivid and natural as those achieved with Editable-DeepSC. This is because Editable-DeepSC reduces the data processing procedures and preserves more semantic mutual information, which eventually leads to more satisfying editing results. Table \ref{tab:main_cbr} shows the compression effectiveness of different methods in terms of CBR defined in (\ref{eq:CBR}). We find that Editable-DeepSC only utilizes around $12.5\%$ or $26.7\%$ of the baselines' CBR, yet it still achieves extraordinary editing effects and outperforms all the baselines. Editable-DeepSC not only performs high-quality editings, but also considerably saves the transmission bandwidth.

\begin{figure*}[h]
    \centering
    \begin{subfigure}{0.230\linewidth}
        \includegraphics[width=\linewidth]{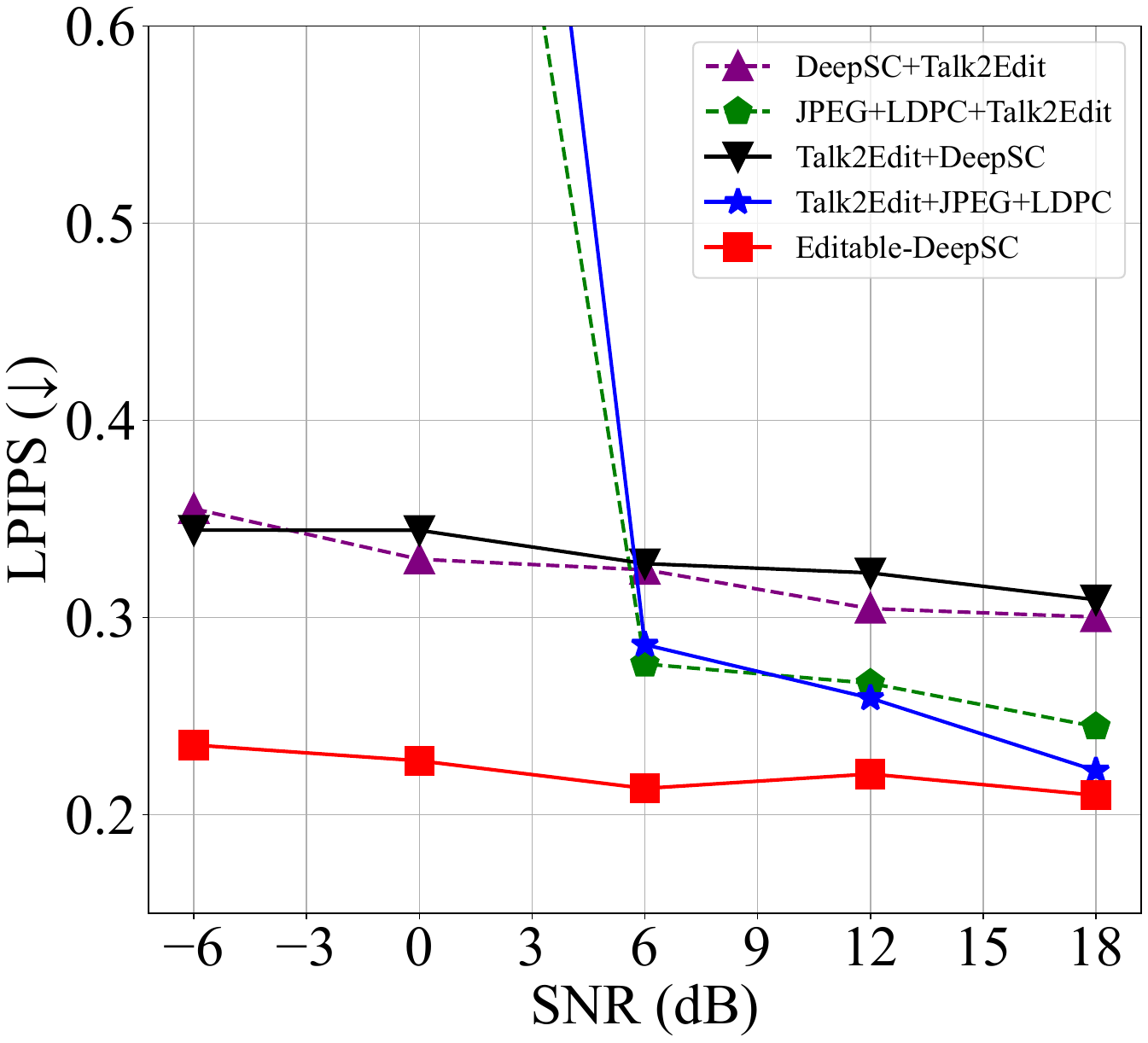}
    \end{subfigure}
    \begin{subfigure}{0.230\linewidth}
        \includegraphics[width=\linewidth]{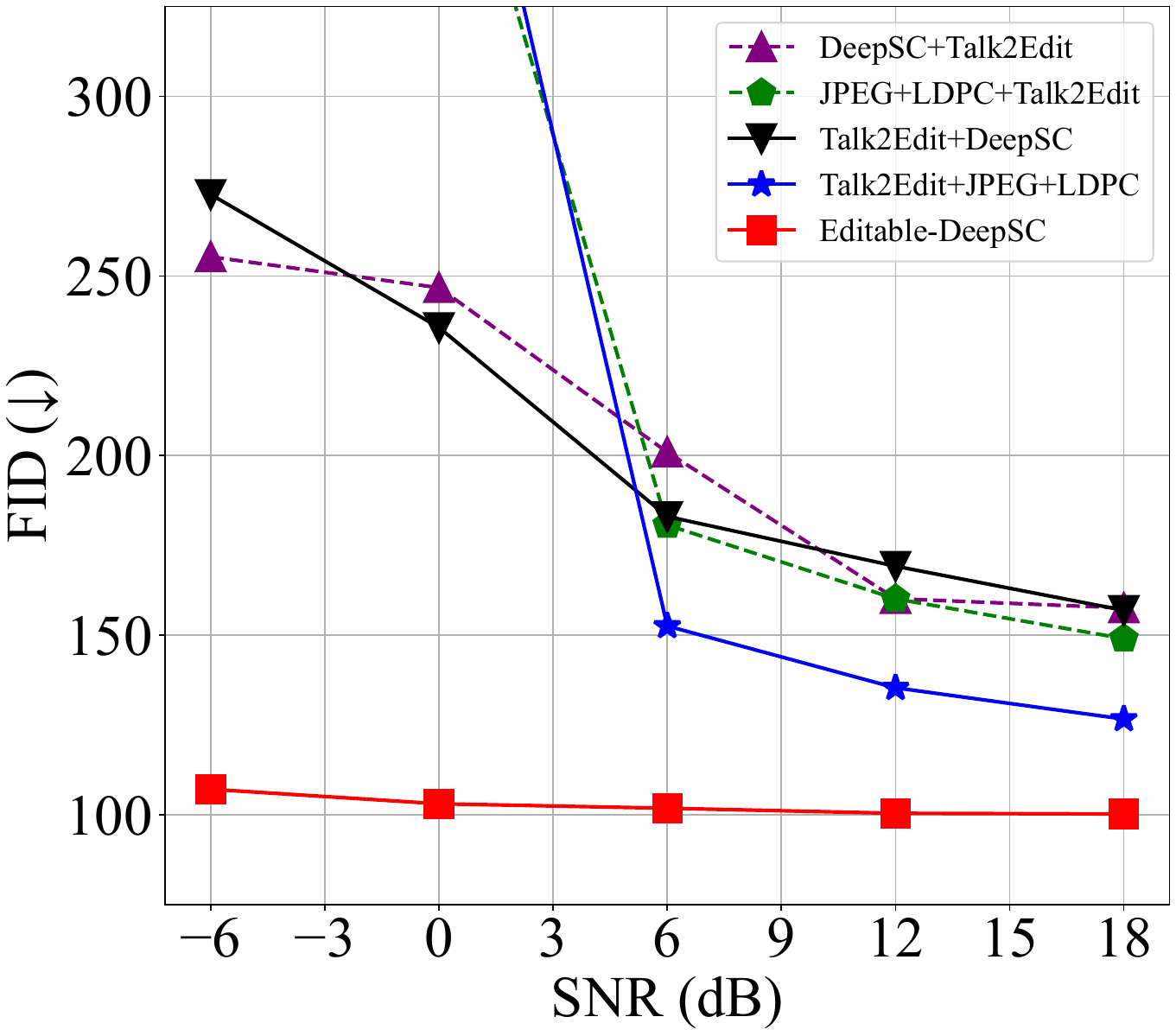}
    \end{subfigure}
    \begin{subfigure}{0.230\linewidth}
        \includegraphics[width=\linewidth]{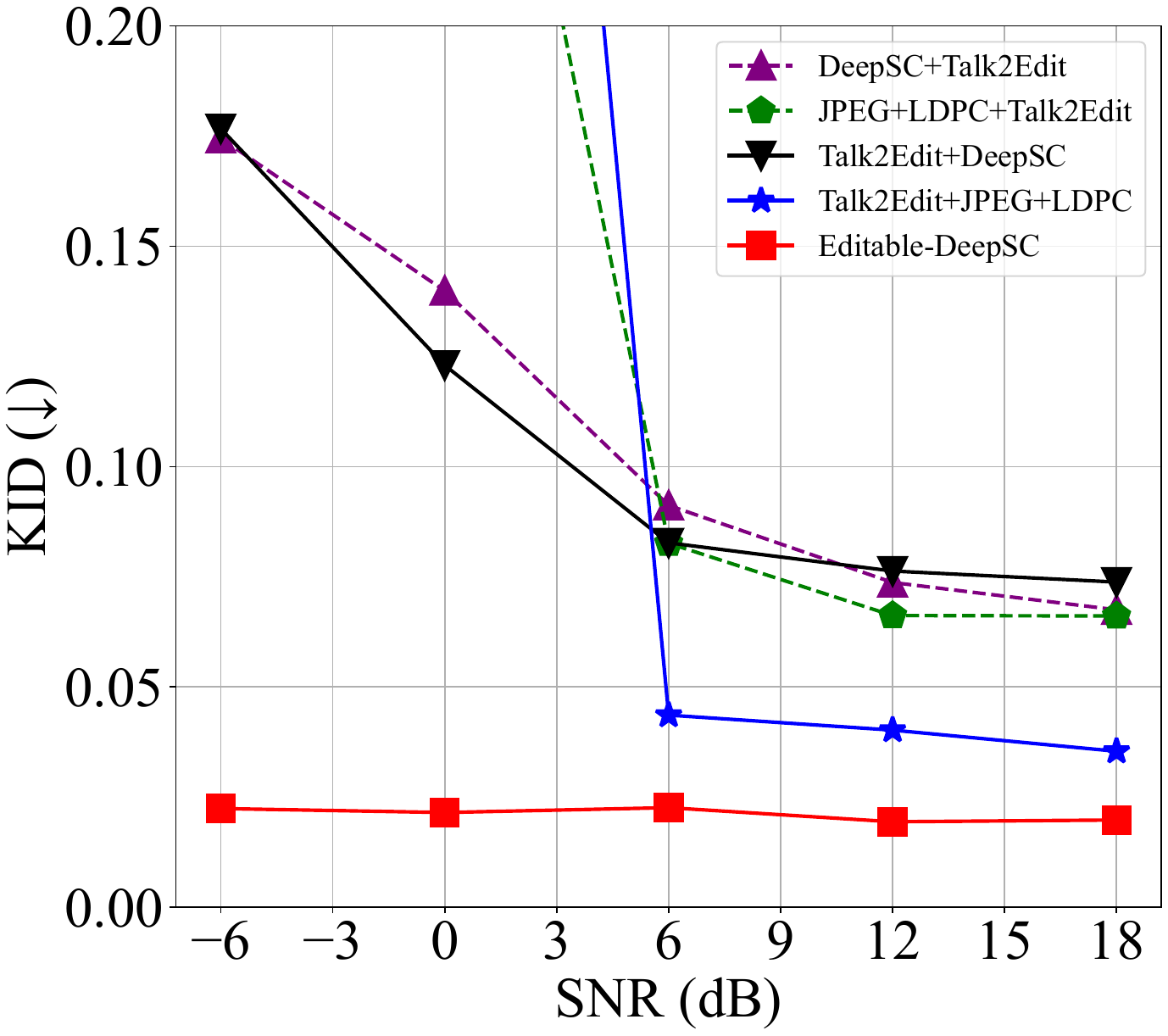}
    \end{subfigure}
    \begin{subfigure}{0.230\linewidth}
        \includegraphics[width=\linewidth]{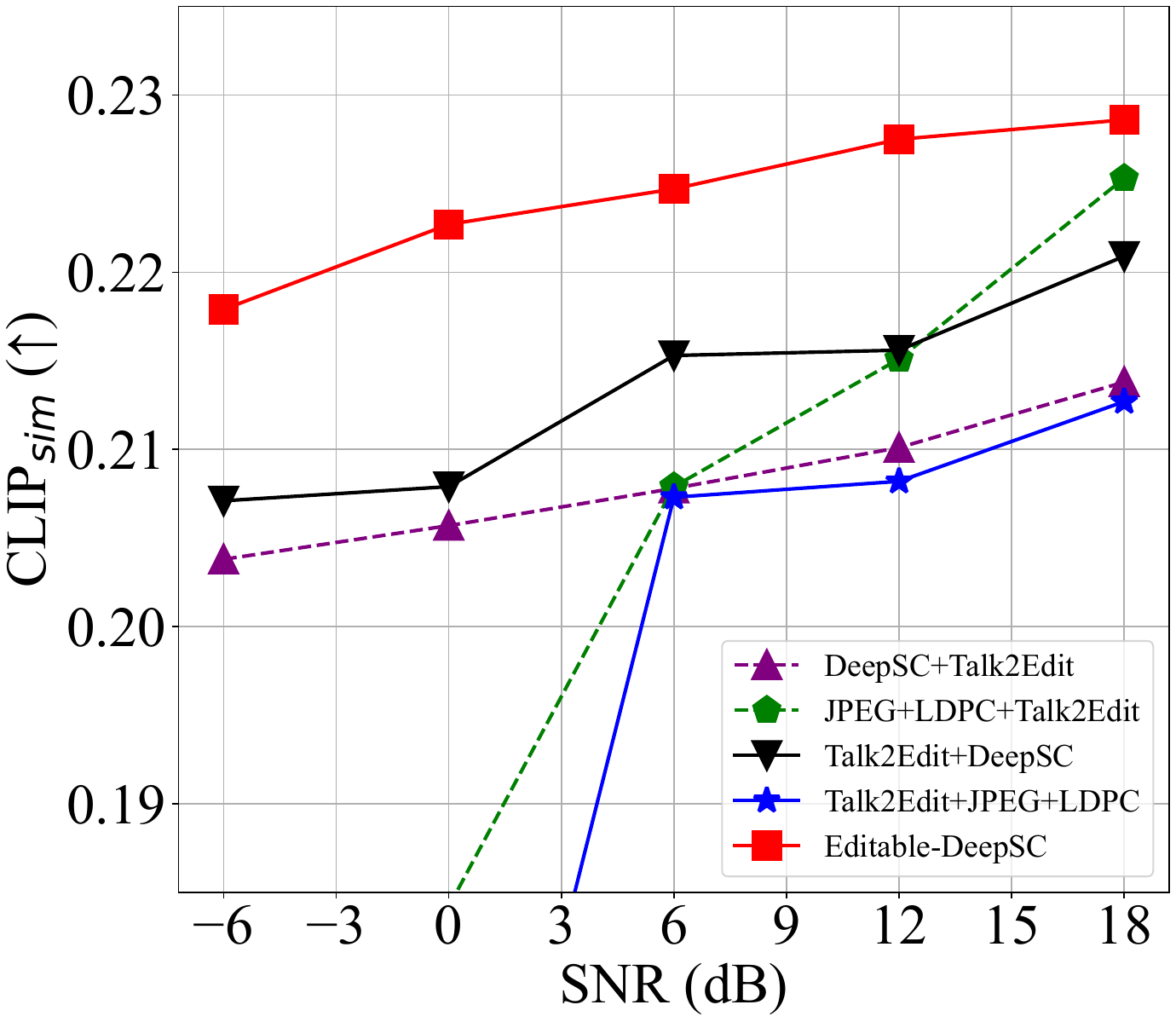}
    \end{subfigure}

    \caption{Quantitative comparison of different methods on the MetFaces dataset (resolution $128 \times 128$) for cross-modal language-driven image editing and transmission tasks. Note that $\downarrow$ indicates that the lower the metric, the better the performance, while $\uparrow$ indicates that the higher the metric, the better the performance.}
    \label{fig:more_metfaces}
\end{figure*}

\begin{figure*}[h]

        \centering
        \begin{subfigure}{0.92\linewidth}
        \centering
        \begin{minipage}[t]{0.140\linewidth}
        \centering
        \includegraphics[width=2.426cm]{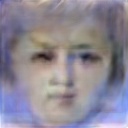}
        \centering
        \end{minipage}
        \begin{minipage}[t]{0.140\linewidth}
        \centering
        \includegraphics[width=2.426cm]{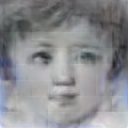}
        \centering
        \end{minipage}
        \begin{minipage}[t]{0.140\linewidth}
        \centering
        \includegraphics[width=2.426cm]{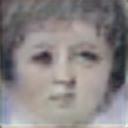}
        \centering
        \end{minipage}
        \begin{minipage}[t]{0.140\linewidth}
        \centering
        \includegraphics[width=2.426cm]{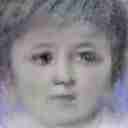}
        \centering
        \end{minipage}
        \begin{minipage}[t]{0.140\linewidth}
        \centering
        \includegraphics[width=2.426cm]{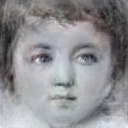}
        \centering
        \end{minipage}
        \begin{minipage}[t]{0.140\linewidth}
        \centering
        \includegraphics[width=2.426cm]{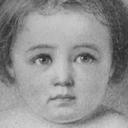}
        \centering
        \end{minipage}
        \end{subfigure}

        \vspace{0.8pt}

        \begin{subfigure}{0.92\linewidth}
        \centering
        \begin{minipage}[t]{0.140\linewidth}
        \centering
        \includegraphics[width=2.426cm]{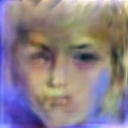}
        \caption*{\footnotesize\textbf{\makecell{DeepSC+Talk2Edit}}}
        \centering
        \end{minipage}
        \begin{minipage}[t]{0.140\linewidth}
        \centering
        \includegraphics[width=2.426cm]{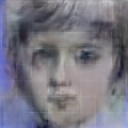}
        \caption*{\footnotesize\textbf{\makecell{JPEG+LDPC\\+Talk2Edit}}}
        \centering
        \end{minipage}
        \begin{minipage}[t]{0.140\linewidth}
        \centering
        \includegraphics[width=2.426cm]{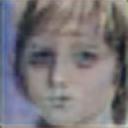}
        \caption*{\footnotesize\textbf{\makecell{Talk2Edit+DeepSC}}}
        \centering
        \end{minipage}
        \begin{minipage}[t]{0.140\linewidth}
        \centering
        \includegraphics[width=2.426cm]{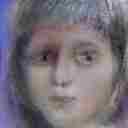}
        \caption*{\footnotesize\textbf{\makecell{Talk2Edit\\+JPEG+LDPC}}}
        \centering
        \end{minipage}
        \begin{minipage}[t]{0.140\linewidth}
        \centering
        \includegraphics[width=2.426cm]{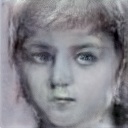}
        \caption*{\footnotesize\textbf{\makecell{Editable-DeepSC}}}
        \centering
        \end{minipage}
        \begin{minipage}[t]{0.140\linewidth}
        \centering
        \includegraphics[width=2.426cm]{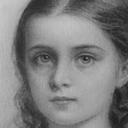}
        \caption*{\footnotesize\textbf{\makecell{Original}}}
        \centering
        \end{minipage}
        \end{subfigure}

    \caption{Qualitative comparison of different methods on the MetFaces dataset (resolution $128 \times 128$) for cross-modal language-driven image editing and transmission tasks at the SNR of $18$ dB. The instructive sentences for the $1$st and $2$nd rows are respectively ``Make the bangs \underline{longer}" and ``What about trying \underline{extremely long} fringe".}
    \label{fig:visualization_more_metfaces}
\end{figure*}

\subsection{High-Resolution Scenario Results}

We then consider more realistic settings where the image resolution is expanded to $1024 \times 1024$ on the CelebA-HQ dataset. Figure \ref{fig:more_celeba_hq} presents the LPIPS, FID, KID, and CLIP$_{sim}$ results while Table \ref{tab:more_cbr} presents the CBR results. The amplification of image pixels objectively increases the complexity of editing and transmission tasks. However, we find that Editable-DeepSC again achieves the best LPIPS scores under almost all the tested cases. As for the FID, KID, and CLIP$_{sim}$ scores, although Editable-DeepSC can't obtain the best results in some cases, it only uses around $0.2\%$ or $1.8\%$ of the baselines' CBR with much lower bandwidth costs as shown in Table \ref{tab:more_cbr}. In general, Editable-DeepSC still exhibits great robustness and efficiency under high-resolution scenarios.

\begin{table}
    \caption{Compression effectiveness of different methods with the resolution of $1024 \times 1024$, measured by the Channel Bandwidth Ratio (CBR) defined in (\ref{eq:CBR}).}
    \label{tab:more_cbr}
    \centering
    \adjustbox{max width=0.5\textwidth}{\begin{tabular}{|c|c|c|}
    \hline
    Method & Resolution & CBR ($\downarrow$) \\
    \hline
    DeepSC+Talk2Edit & $1024 \times 1024$ & 0.0833 \\
    JPEG+LDPC+Talk2Edit & $1024 \times 1024$ & 0.0111 \\
    Talk2Edit+DeepSC & $1024 \times 1024$ & 0.0833 \\
    Talk2Edit+JPEG+LDPC & $1024 \times 1024$ & 0.0111 \\
    \gc{Editable-DeepSC} & \gc$1024 \times 1024$ & \gc\textbf{0.0002} \\
    \hline
    \end{tabular}}
\end{table}

Figure \ref{fig:visualization_more_celeba_hq} shows the visualization results of different methods on the CelebA-HQ dataset at the SNR of $6$ dB. We notice that Editable-DeepSC can acquire better editing effects than the other methods, with fewer artifacts or blurs. In particular, for the $2$nd row of images where the textual instruction is to make the face slightly younger, Editable-DeepSC manages to deal with this request in a fine-grained way and moderately alter the degree of youth. However, all the other methods overly modify the original face to appear too immature and there is the problem of excessive editing. This again validates the necessity to integrate editings into the communication chain for reduced data processing procedures, which can retain more semantic mutual information from the original images and precisely conduct fine-grained editings.

\subsection{Out-Of-Distribution (OOD) Scenario Results}

We also evaluate different methods under out-of-distribution (OOD) settings on the MetFaces dataset, where the distributions of training data and testing data are mismatched. From Figure \ref{fig:more_metfaces}, we observe that Editable-DeepSC continuously obtains the best LPIPS, FID, KID, and CLIP$_{sim}$ scores. These results demonstrate that Editable-DeepSC possesses outstanding OOD generalizability, and is able to consistently surpass those baselines that separately communicate and edit even on unseen artistic datasets.

The visualization results in Figure \ref{fig:visualization_more_metfaces} further demonstrate that Editable-DeepSC can precisely edit the OOD images without influencing other unrelated regions. In contrast, all the other methods struggle to realize the semantic editing while maintaining the fidelity, and encounter image quality degradation or even collapse under the more rigorous OOD scenarios. Our combination of editings and communications has indeed come into effect for better tasks performance.

\begin{figure}[t]

        \begin{subfigure}{\linewidth}
        \centering
        \begin{minipage}[t]{0.05\linewidth}
        \centering
        \vspace{-53pt}
        \rotatebox{90}{\textbf{CelebA}}
        \centering
        \end{minipage}
        \begin{minipage}[t]{0.278\linewidth}
        \centering
        \includegraphics[width=2.54cm]{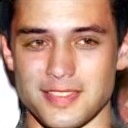}
        \centering
        \end{minipage}
        \begin{minipage}[t]{0.278\linewidth}
        \centering
        \includegraphics[width=2.54cm]{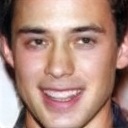}
        \centering
        \end{minipage}
        \begin{minipage}[t]{0.278\linewidth}
        \centering
        \includegraphics[width=2.54cm]{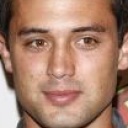}
        \centering
        \end{minipage}
        \end{subfigure}

        \vspace{0.9pt}

        \begin{subfigure}{\linewidth}
        \centering
        \begin{minipage}[t]{0.05\linewidth}
        \centering
        \vspace{-62pt}
        \rotatebox{90}{\textbf{CelebA-HQ}}
        \centering
        \end{minipage}
        \begin{minipage}[t]{0.278\linewidth}
        \centering
        \includegraphics[width=2.54cm]{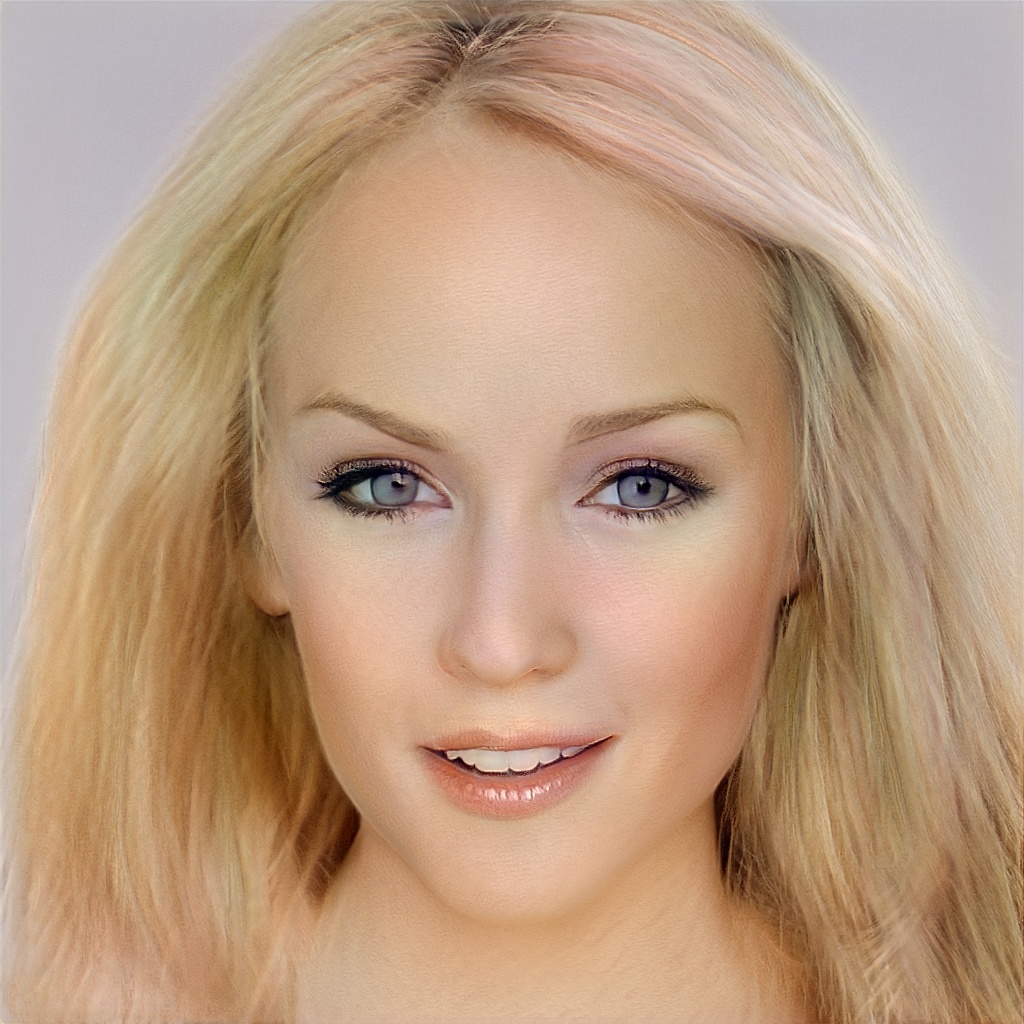}
        \centering
        \end{minipage}
        \begin{minipage}[t]{0.278\linewidth}
        \centering
        \includegraphics[width=2.54cm]{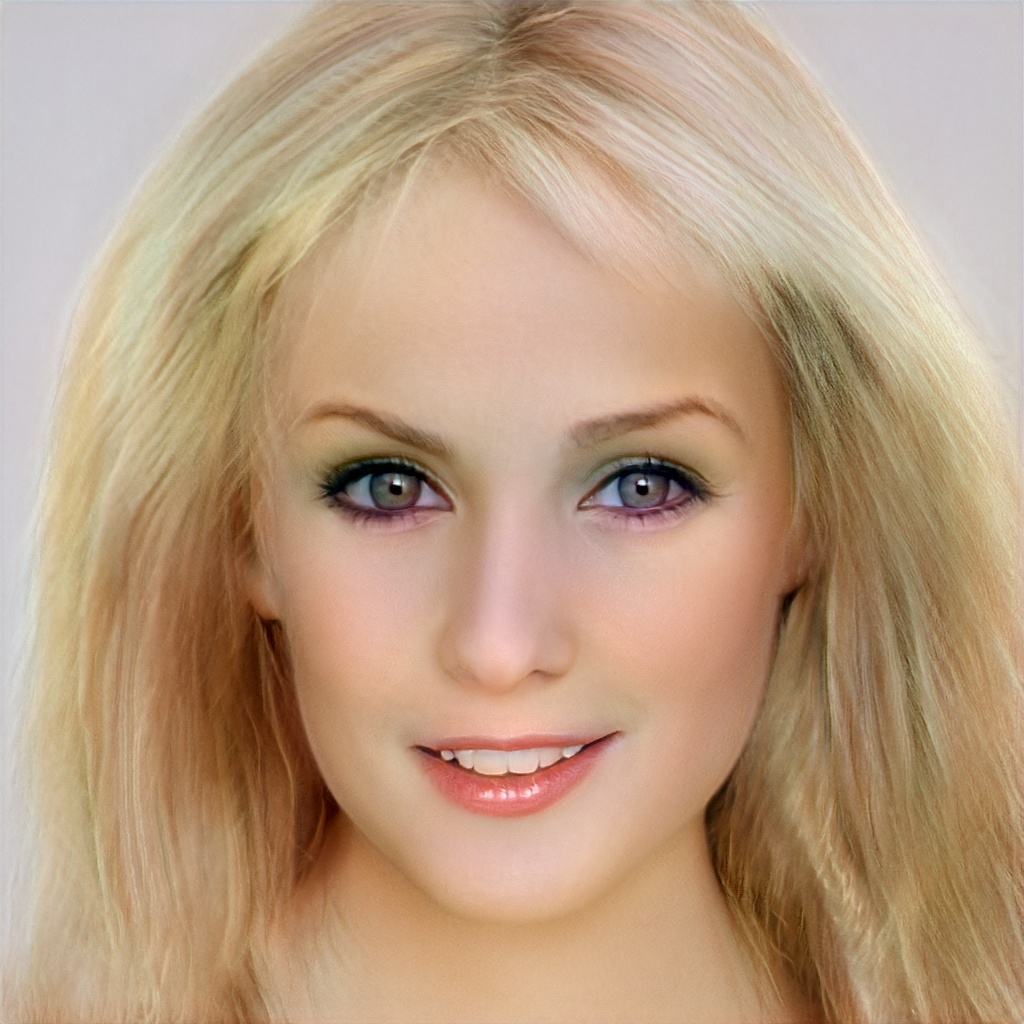}
        \centering
        \end{minipage}
        \begin{minipage}[t]{0.278\linewidth}
        \centering
        \includegraphics[width=2.54cm]{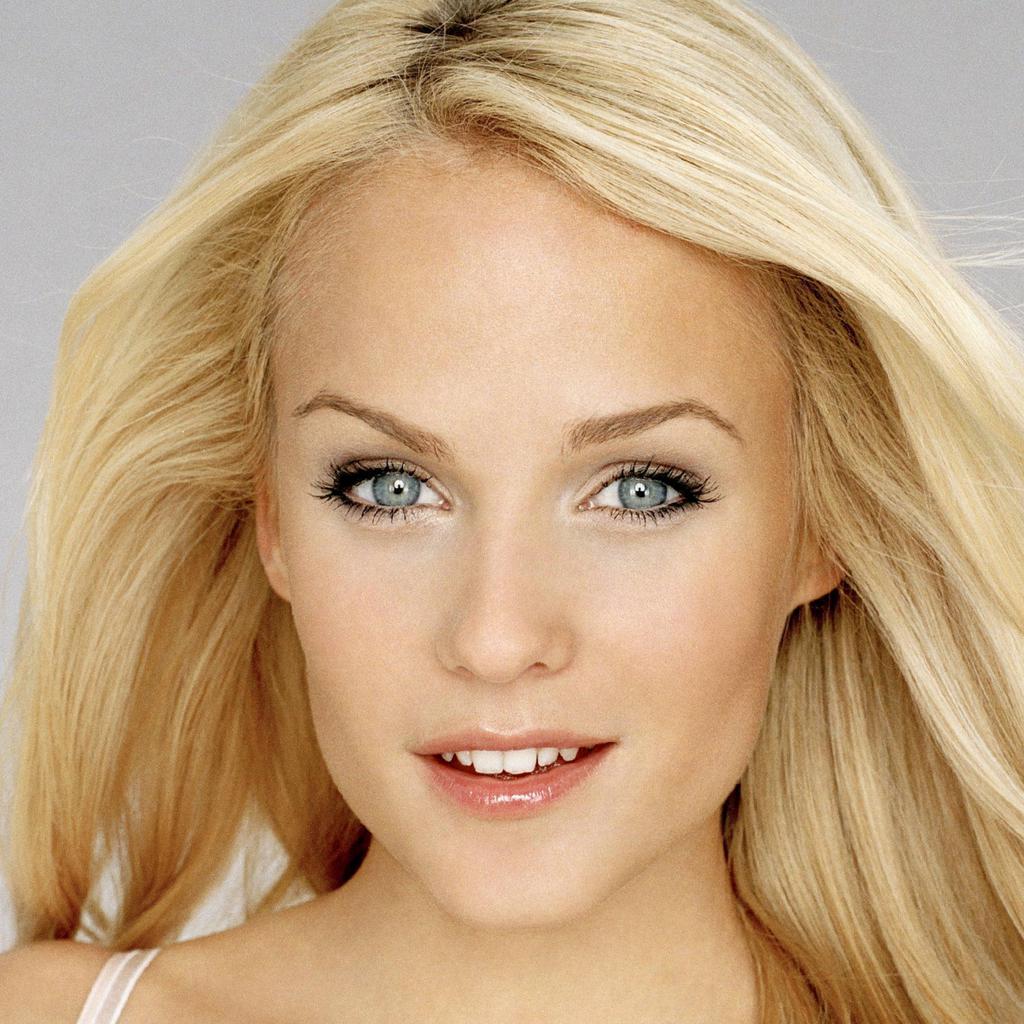}
        \centering
        \end{minipage}
        \end{subfigure}

        \vspace{0.9pt}

        \begin{subfigure}{\linewidth}
        \centering
        \begin{minipage}[t]{0.05\linewidth}
        \centering
        \vspace{-57pt}
        \rotatebox{90}{\textbf{MetFaces}}
        \centering
        \end{minipage}
        \begin{minipage}[t]{0.278\linewidth}
        \centering
        \includegraphics[width=2.54cm]{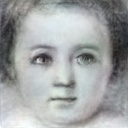}
        \caption*{\footnotesize\textbf{\makecell{w/o IAM}}}
        \centering
        \end{minipage}
        \begin{minipage}[t]{0.278\linewidth}
        \centering
        \includegraphics[width=2.54cm]{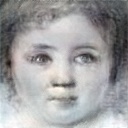}
        \caption*{\footnotesize\textbf{\makecell{w/ IAM}}}
        \centering
        \end{minipage}
        \begin{minipage}[t]{0.278\linewidth}
        \centering
        \includegraphics[width=2.54cm]{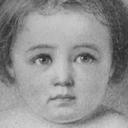}
        \caption*{\footnotesize\textbf{\makecell{Original}}}
        \centering
        \end{minipage}
        \end{subfigure}

    \caption{Qualitative results about whether the iterative attributes matching (IAM) mechanism is incorporated at the SNR of $-6$ dB. The instructive sentences for the $1$st, $2$nd, and $3$rd rows are ``Smile \underline{more}", ``What about trying bangs that leaves \underline{half of} the forehead visible", and ``Make the bangs \underline{longer}".}
    \label{fig:ablation_iam}
\end{figure}

\begin{figure}[t]

        \begin{subfigure}{\linewidth}
        \centering
        \begin{minipage}[t]{0.05\linewidth}
        \centering
        \vspace{-53pt}
        \rotatebox{90}{\textbf{CelebA}}
        \centering
        \end{minipage}
        \begin{minipage}[t]{0.278\linewidth}
        \centering
        \includegraphics[width=2.54cm]{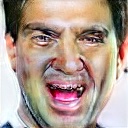}
        \centering
        \end{minipage}
        \begin{minipage}[t]{0.278\linewidth}
        \centering
        \includegraphics[width=2.54cm]{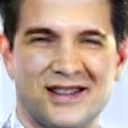}
        \centering
        \end{minipage}
        \begin{minipage}[t]{0.278\linewidth}
        \centering
        \includegraphics[width=2.54cm]{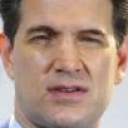}
        \centering
        \end{minipage}
        \end{subfigure}

        \vspace{0.9pt}

        \begin{subfigure}{\linewidth}
        \centering
        \begin{minipage}[t]{0.05\linewidth}
        \centering
        \vspace{-62pt}
        \rotatebox{90}{\textbf{CelebA-HQ}}
        \centering
        \end{minipage}
        \begin{minipage}[t]{0.278\linewidth}
        \centering
        \includegraphics[width=2.54cm]{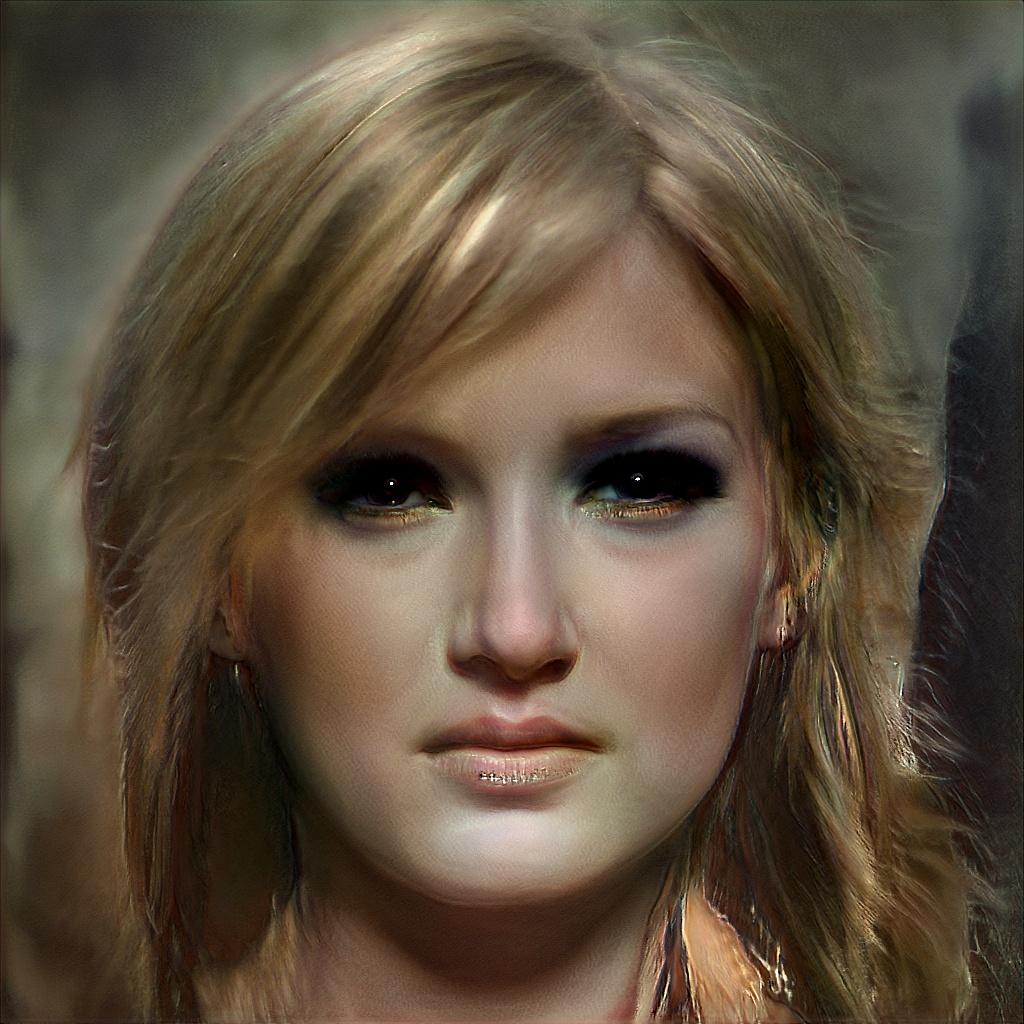}
        \centering
        \end{minipage}
        \begin{minipage}[t]{0.278\linewidth}
        \centering
        \includegraphics[width=2.54cm]{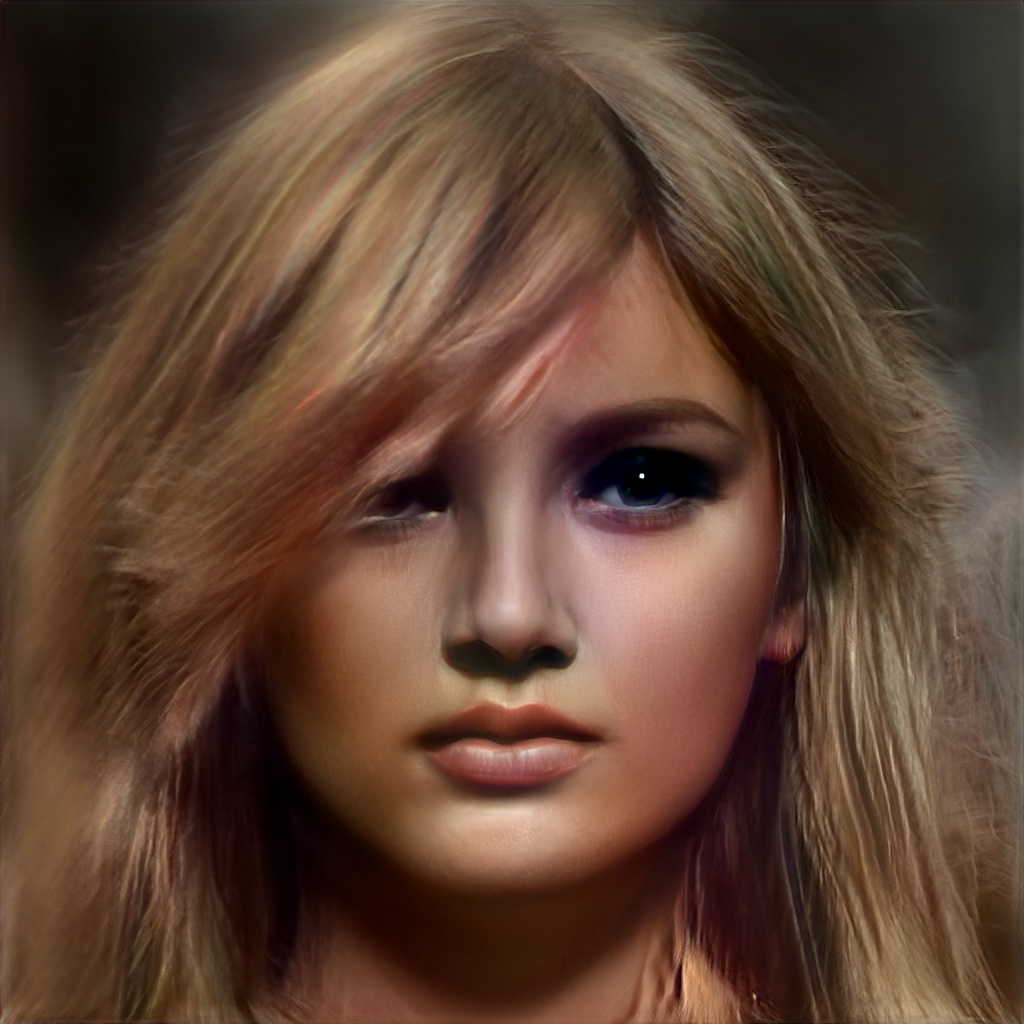}
        \centering
        \end{minipage}
        \begin{minipage}[t]{0.278\linewidth}
        \centering
        \includegraphics[width=2.54cm]{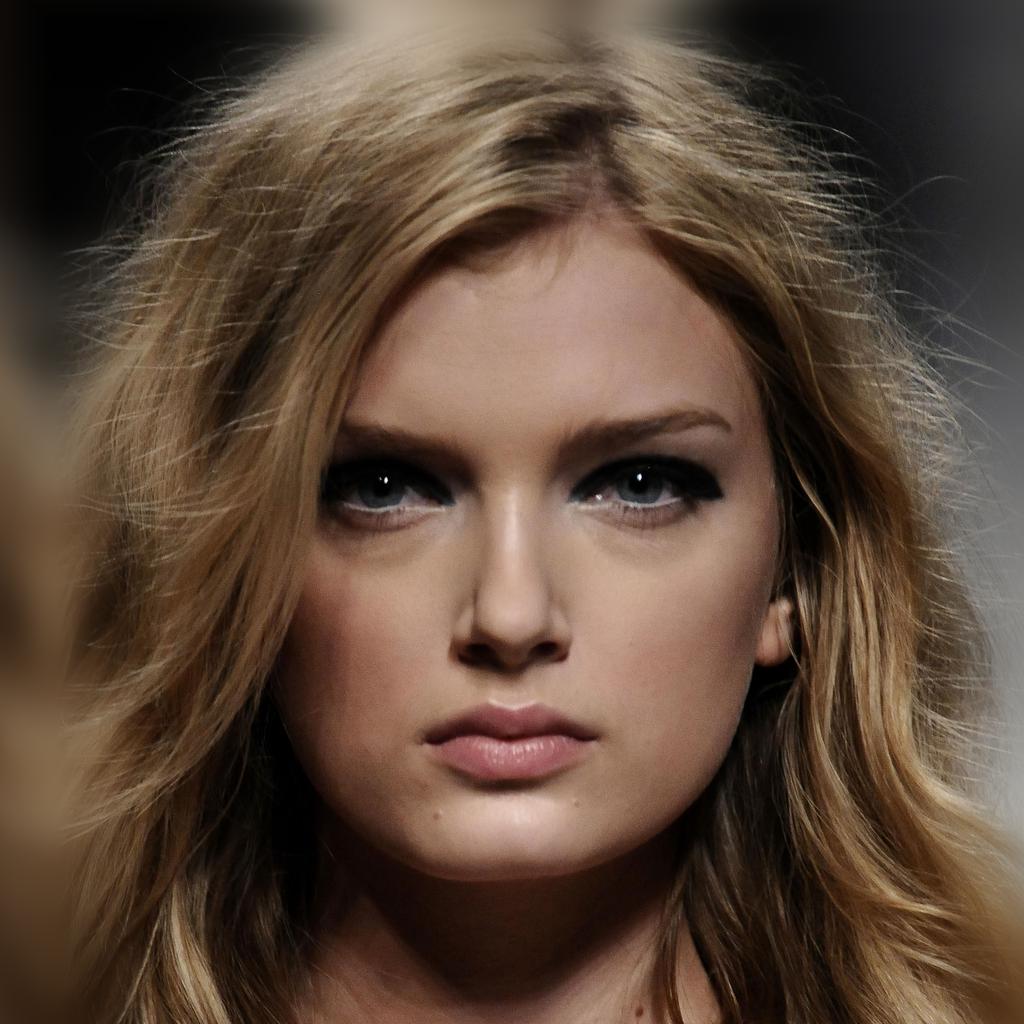}
        \centering
        \end{minipage}
        \end{subfigure}

        \vspace{0.9pt}

        \begin{subfigure}{\linewidth}
        \centering
        \begin{minipage}[t]{0.05\linewidth}
        \centering
        \vspace{-57pt}
        \rotatebox{90}{\textbf{MetFaces}}
        \centering
        \end{minipage}
        \begin{minipage}[t]{0.278\linewidth}
        \centering
        \includegraphics[width=2.54cm]{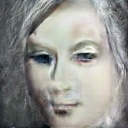}
        \caption*{\footnotesize\textbf{\makecell{w/o FT}}}
        \centering
        \end{minipage}
        \begin{minipage}[t]{0.278\linewidth}
        \centering
        \includegraphics[width=2.54cm]{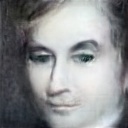}
        \caption*{\footnotesize\textbf{\makecell{w/ FT}}}
        \centering
        \end{minipage}
        \begin{minipage}[t]{0.278\linewidth}
        \centering
        \includegraphics[width=2.54cm]{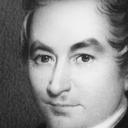}
        \caption*{\footnotesize\textbf{\makecell{Original}}}
        \centering
        \end{minipage}
        \end{subfigure}

    \caption{Qualitative results about whether the fine-tuning (FT) mechanism is incorporated at the SNR of $-6$ dB. The instructive sentences for the $1$st, $2$nd, and $3$rd rows are ``Smile \underline{more}", ``Add \underline{long} bangs", and ``Make the fringe just \underline{a little longer}".}
    \label{fig:ablation_ft}
\end{figure}

\subsection{Further Analysis}
\label{sec:further_analysis}

\begin{table*}[t]
    \caption{Quantitative results under different values of $r_{inv}$ and $r_{ft}$ on the CelebA dataset (resolution $128 \times 128$) at the SNR of $-6$ dB.}
    \label{tab:ablation_gan_inversion_ft_another}
    \centering
    \adjustbox{max width=\linewidth}{
    \begin{tabular}{|c|c|c|c|c|c|c|c|c|c|c|}
    \hline
    \multicolumn{1}{|c|}{\multirow{2}{*}{Metric}} & \multicolumn{5}{c|}{$r_{inv}$} & \multicolumn{5}{c|}{$r_{ft}$} \\
    \cline{2-11}
    \multicolumn{1}{|c|}{} & 70 & 90 & \gc{110} & 130 & 150 & 2 & 4 & \gc{6} & 8 & 10 \\
    \hline
    LPIPS ($\downarrow$) & 0.1769 & 0.1633 & \gc\textbf{0.1616} & 0.1642 & 0.1624 & 0.2110 & 0.1705 & \gc\textbf{0.1616} & 0.1643 & 0.1645 \\
    FID ($\downarrow$) & 67.02 & 66.99 & \gc\textbf{66.64} & 68.75 & 71.12 & 84.24 & 71.39 & \gc\textbf{66.64} & 69.21 & 70.63 \\
    KID ($\downarrow$) & 0.0018 & 0.0009 & \gc\textbf{-0.0012} & -0.0005 & -0.0010 & 0.0132 & 0.0004 & \gc\textbf{-0.0012} & -0.0007 & -0.0003 \\
    CLIP$_{sim}$ ($\uparrow$) & 0.2296 & 0.2321 & \gc\textbf{0.2439} & 0.2367 & 0.2358 & 0.2266 & 0.2364 & \gc\textbf{0.2439} & 0.2310 & 0.2336 \\
    GPU Energy Costs ($\downarrow$) & \textbf{252.19 mWh} & 308.19 mWh & \gc362.05 mWh & 419.49 mWh & 470.95 mWh & \textbf{358.26} mWh & 360.74 mWh & \gc362.05 mWh & 363.61 mWh & 367.26 mWh \\
    \hline
    \end{tabular}}
\end{table*}

\begin{table}[t]
    \begin{minipage}{0.48\textwidth}
    \centering
    \caption{Quantitative results about whether the fine-tuning (FT) mechanism is incorporated at the SNR of $-6$ dB.}
    \label{tab:ablation_ft}
    \centering
    \adjustbox{max width=\linewidth}{
    \begin{tabular}{|c|c|c|c|c|c|}
    \hline
    Dataset & Method & LPIPS ($\downarrow$) & FID ($\downarrow$) & KID ($\downarrow$) & CLIP$_{sim}$ ($\uparrow$) \\
    \hline
    \multirow{2}[0]{*}{CelebA} & w/o FT & 0.3105 & 96.73 & 0.0292 & 0.2213 \\
    & \gc{w/ FT} & \gc\textbf{0.1616} & \gc\textbf{66.64} & \gc\textbf{-0.0012} & \gc\textbf{0.2439} \\
    \hline
    \multirow{2}[0]{*}{CelebA-HQ} & w/o FT & 0.4790 & 100.03 & 0.0209 & 0.2022 \\
    & \gc{w/ FT} & \gc\textbf{0.3602} & \gc\textbf{88.47} & \gc\textbf{0.0148} & \gc\textbf{0.2141} \\
    \hline
    \multirow{2}[0]{*}{MetFaces} & w/o FT & 0.3501 & 141.63 & 0.0821 & 0.2023 \\
    & \gc{w/ FT} & \gc\textbf{0.2354} & \gc\textbf{107.12} & \gc\textbf{0.0224} & \gc\textbf{0.2179} \\
    \hline
    \end{tabular}}
    \end{minipage}
    \hfill
\end{table}

\textbf{The influences of pre-trained GAN inversion.} First, we explore the impacts of the proposed GAN inversion based semantic coding by changing the value of $r_{inv}$. From Table \ref{tab:ablation_gan_inversion_ft_another}, we notice that when $r_{inv}$ is small, increasing its value indeed improves the editing performance as the GAN inversion is more fully optimized, which validates the contribution of this component. However, when $r_{inv}$ is large, further increasing its value cannot improve the editing performance and will also introduce more GPU energy costs. Thus, our adoption of $r_{inv} = 110$ is reasonable, for it strikes a better balance between editing effects and GPU energy costs.

\textbf{The influences of iterative attributes matching (IAM).} Next, we study the impacts of the iterative attributes matching mechanism by removing it from Editable-DeepSC. As shown in Figure \ref{fig:ablation_iam}, when the iterative attributes matching mechanism is excluded, Editable-DeepSC will be unable to precisely perform the anticipated editings according to the given texts in a fine-grained way. This again provides evidence for the contribution of our IAM mechanism.

\textbf{The influences of model fine-tuning.} Similarly, we also investigate the impacts of the model fine-tuning mechanism in Table \ref{tab:ablation_gan_inversion_ft_another}, Table \ref{tab:ablation_ft}, and Figure \ref{fig:ablation_ft}. We discover that under the extreme channel condition of $-6$ dB SNR, the fine-tuning mechanism plays a crucial role in preserving the quality and fidelity of edited images, while removing it will conversely generate unnatural or even collapsed regions. These results also verify the contribution of our fine-tuning mechanism. Moreover, our adoption of $r_{ft} = 6$ yields a satisfactory trade-off between fine-tuning effects and GPU energy costs.

\begin{table}[t]
    \begin{minipage}{0.48\textwidth}
    \centering
    \caption{GPU energy costs ($\downarrow$) of different methods averaged over each image under different resolutions.}
    \label{tab:energy_analysis}
    \centering
    \adjustbox{max width=\linewidth}{
    \begin{tabular}{|c|c|c|}
    \hline
    Method & $128 \times 128$ & $1024 \times 1024$ \\
    \hline
    DeepSC+Talk2Edit & 1432.94 mWh & 3547.15 mWh \\
    JPEG+LDPC+Talk2Edit & 1540.53 mWh & 3732.17 mWh \\
    Talk2Edit+DeepSC & 1462.73 mWh & 3722.05 mWh \\
    Talk2Edit+JPEG+LDPC & 1485.72 mWh & 3925.87 mWh \\
    \gc{Editable-DeepSC} & \gc\textbf{354.24 mWh} & \gc\textbf{942.44 mWh} \\
    \hline
    \end{tabular}}
    \end{minipage}
\end{table}

\begin{table}[t]
    \centering
    \caption{The numbers of parameters in various parts of Editable-DeepSC. We also report the ratio of tuned parameters.}
    \label{tab:analysis_number_parameter}
    \adjustbox{max width=\linewidth}{\begin{tabular}{|c|c|c|c|}
    \hline
    Frozen Parameters & Tuned Parameters & Total Parameters & \gc{Tuned Ratio} \\
    \hline
    $7.70\times10^7$ & $0.21\times10^7$ & $7.91\times10^7$ & \gc$\mathbf{2.65\%}$ \\
    \hline
    \end{tabular}}
\end{table}

\begin{table}[h]
    \centering
    \caption{Component contribution analysis under varying channel conditions. We report the CLIP$_{sim}$ ($\uparrow$) values of different model variants on the CelebA dataset.}
    \label{tab:component_contribution}
    \adjustbox{max width=\linewidth}{
    \begin{tabular}{|c|c|c|c|c|c|}
    \hline
    Model Variant & $-6$ dB & $0$ dB & $6$ dB & $12$ dB & $18$ dB \\
    \hline
    \gc{Editable-DeepSC} & \gc\textbf{0.2439} & \gc\textbf{0.2454} & \gc\textbf{0.2458} & \gc\textbf{0.2476} & \gc\textbf{0.2487} \\
    w/ weak GAN inversion & 0.2296 & 0.2329 & 0.2357 & 0.2371 & 0.2388 \\
    w/o IAM & 0.2266 & 0.2304 & 0.2312 & 0.2297 & 0.2310 \\
    w/o FT & 0.2213 & 0.2242 & 0.2318 & 0.2362 & 0.2366 \\
    \hline
    \end{tabular}}
\end{table}

\textbf{Computational costs.} Next, we provide analysis on computational costs. From Table \ref{tab:energy_analysis}, Editable-DeepSC has the smallest GPU energy costs and surpasses all the baselines. This is because Editable-DeepSC jointly optimizes the communications and editings with reduced data processing procedures, while the separate baselines repeatedly encode the data and result in low efficiency. Furthermore, Editable-DeepSC consumes less than 1600 MiB GPU memory for the resolution of $128\times128$ and less than 3000 MiB GPU memory for the resolution of $1024\times1024$. Thus, Editable-DeepSC achieves a good trade-off between utility and efficiency.

\textbf{Parameters analysis.} From Table \ref{tab:analysis_number_parameter}, we note that the fine-tuned parameters only take up $2.65\%$ of the total, while the remaining $97.35\%$ of the total parameters are kept frozen. Nonetheless, the fine-tuning mechanism still results in significant performance gains. This further confirms the practicality and effectiveness of our fine-tuning method.

\textbf{Component contribution analysis under varying channel conditions.} Table \ref{tab:component_contribution} quantitatively analyzes the contribution of each key component under varying channel conditions. The full Editable-DeepSC model consistently achieves the highest CLIP$_{sim}$ across all SNR levels, demonstrating the synergistic effect of its integrated design. The variant with weakened GAN inversion (adopting $r_{inv} = 70$) shows a consistent performance drop, indicating that high-fidelity semantic compression is foundational for preserving editing semantics. The most significant performance degradation, especially at medium-to-high SNRs, occurs when removing the Iterative Attributes Matching (IAM) mechanism, highlighting its critical role in achieving precise semantic alignment with text instructions. Conversely, the variant without fine-tuning (FT) suffers the most severe performance loss at low SNRs (e.g., $-6$ dB), underscoring that the SNR-aware adaptation module is primarily responsible for maintaining robustness in harsh channel conditions. In summary, GAN inversion provides the essential compressed representation, the JECC mechanism enables accurate editing, and fine-tuning ensures channel resilience. Their combination in the full Editable-DeepSC model optimally balances these requirements for reliable cross-modal semantic communications.


\section{Conclusion}
\label{conclusion}

In this paper, we propose Editable-DeepSC, a novel cross-modal semantic communication approach that tackles the communication challenges under dynamic facial editing scenarios. Specifically, we first theoretically analyze different transmission schemes that separately handle communications and editings. We find that they actually increase the data processing procedures and are more likely to lose semantic mutual information. In light of this, we emphasize the necessity to integrate editings into the communication chain, jointly optimize these two aspects, and preserve more semantic mutual information. To compress the high-dimensional data, we leverage inversion methods based on pre-trained StyleGAN priors for semantic coding. To tackle the dynamic channel conditions, we propose SNR-aware channel coding based on model fine-tuning, which introduces two lightweight trainable adapters to capture the distribution of new conditions. Extensive experiments demonstrate that compared to existing methods, Editable-DeepSC exhibits superior editing effects while significantly saving the bandwidth under multiple settings.

\bibliographystyle{IEEEtran}
\bibliography{refs}

\end{document}